\documentclass[preprint]{elsarticle}

\usepackage[]{todonotes}
\usepackage{graphicx}

\usepackage{listings}
\usepackage{xspace}
\usepackage{hyperref}
\usepackage{booktabs}
\usepackage{graphicx}
\usepackage[htt]{hyphenat}
\usepackage[caption=false]{subfig}

\usepackage[utf8]{inputenc} %
\usepackage{pmboxdraw} %

\usepackage{tikz} %

\newcommand\YAMLkeystyle{\color{black}\bfseries}
\newcommand\YAMLvaluestyle{\color{blue}\mdseries}

\makeatletter

\let\old@sverb\@sverb
\def\@sverb#1{\old@sverb{#1}\zz}
\def\zz#1{#1\ifx\@undefined#1\else\penalty\z@\expandafter\zz\fi}

\newcommand\language@yaml{yaml}

\expandafter\expandafter\expandafter\lstdefinelanguage
\expandafter{\language@yaml}
{
  keywords={true,false,null,y,n},
  keywordstyle=\color{darkgray}\bfseries,
  basicstyle=\YAMLkeystyle,                                 %
  sensitive=false,
  comment=[l]{\#},
  morecomment=[s]{/*}{*/},
  commentstyle=\color{purple}\ttfamily,
  stringstyle=\YAMLvaluestyle\ttfamily,
  moredelim=[l][\color{orange}]{\&},
  moredelim=[l][\color{magenta}]{*},
  morestring=[b]',
  morestring=[b]",
  literate =    {---}{{\ProcessThreeDashes}}3
                {\ -\ }{{\mdseries\ -\ }}3,
}

\newcommand{\etal}{et al.\xspace}
\newcommand{\ie}{i.e.,\xspace}
\newcommand{\eg}{e.g.,\xspace}
\newcommand{\fig}[1]{Fig.~\ref{#1}}

\newcommand{\tab}[1]{Table~\ref{#1}}
\newcommand{\sect}[1]{Section~\ref{#1}}

\newcommand{\github}{GitHub\xspace}

\newcommand{\git}{git\xspace}

\newcommand{\gha}{GHA\xspace}
\newcommand{\actions}{GitHub Actions\xspace}
\newcommand{\travis}{Travis\xspace}
\newcommand{\circleci}{CircleCI\xspace}

\newcommand{\appveyor}{AppVeyor\xspace}
\newcommand{\gitlabci}{GitLab CI/CD\xspace}

\newcommand{\jenkins}{Jenkins\xspace}

\newcommand{\gawd}{\emph{gawd}\xspace}
\newcommand{\figsize}{0.9\textwidth}
\newcommand{\yaml}{YAML\xspace}
\newcommand{\entity}[1]{\textsf{#1}\xspace}

\usepackage[flushleft]{threeparttable}

\usepackage[framemethod=TikZ]{mdframed}
\usepackage{color, colortbl}
\usepackage{pifont}
\definecolor{gray}{cmyk}{0,0,0,0.50}
\usepackage{multirow}
\usepackage{diagbox}

\global\mdfdefinestyle{exampledefault}{linewidth=1pt, linecolor=lightgray!80, backgroundcolor=lightgray!20, innerleftmargin=6pt, innerrightmargin=6pt, innertopmargin=4pt, innerbottommargin=4pt, nobreak=true,roundcorner=5pt}
\newenvironment{custombox}{\begin{mdframed}[style=exampledefault]\emph{Takeaways.}}{\end{mdframed}\smallskip}

\def\BibTeX{{\rm B\kern-.05em{\sc i\kern-.025em b}\kern-.08em
    T\kern-.1667em\lower.7ex\hbox{E}\kern-.125emX}}

\newcommand{\changed}[1]{#1}
\newcommand{\changedTo}[1]{#1}

\journal{Journal of Systems \& Software}
\begin{document}

\begin{frontmatter}
\title{An Empirical Study of the Evolution\\ of GitHub Actions Workflows}

\author[1]{Pooya Rostami Mazrae}
\ead{pooya.rostamimazrae@umons.ac.be}
\author[1]{Alexandre Decan\fnref{fnrs}}
\ead{alexandre.decan@umons.ac.be}
\author[1]{Tom Mens}
\ead{tom.mens@umons.ac.be}
\author[2]{Mairieli Wessel}
\ead{mairieli.wessel@ru.nl}

\fntext[fnrs]{F.R.S.-FNRS Research Associate}

\affiliation[1]{organization={Software Engineering Lab, University of Mons},
        city={Mons},
        country={Belgium}}

\affiliation[2]{organization={Radboud University},
        city={Nijmegen},
        country={The Netherlands}}

\begin{abstract}

CI/CD practices play a significant role during collaborative software development by automating time-consuming and repetitive tasks such as testing, building, quality checking, dependency and security management.
GitHub Actions, the CI/CD tool integrated into GitHub, allows repository maintainers to automate development workflows.
We conducted a mixed methods analysis of GitHub Actions workflow changes over time.
Through a \changed{preliminary} qualitative analysis of 439 modified workflow files we identified seven types of conceptual changes to workflows.
Next, we performed a quantitative analysis over \changed{49}K+ GitHub repositories totaling \changed{267}K+ %
workflow change histories and \changed{3.4}M+ workflow file versions from November 2019 to \changed{August 2025}.
This analysis revealed that repositories contain a median of \changed{three}  workflow files, and \changed{7.3}\% of all workflow files are being changed every week.
The changes made to workflows tend to be small, with about three-quarters containing only a single change.
The large majority of the observed changes have to do with task configuration and task specification in workflow jobs.
\changedTo{We did not find any conclusive evidence of the effect of LLM coding tools or other major technological changes on workflow creation and workflow maintenance frequency.
Our findings highlight the need for improved tooling to support fine-grained maintenance tasks, such as a broader adoption of dependency management and AI-based support for ensuring and sustaining workflow security and quality.}

\end{abstract}


\end{frontmatter}

\section{Introduction} \label{sec:introduction}

Continuous Integration and Continuous Delivery (CI/CD) practices have become an essential part of software development~\cite{duvall2007continuous,Shahin2017,Wessel2023-chapter}, helping developers achieve more efficient and less labor-intensive software releases, while maintaining high-quality standards~\cite{savor2016continuous}. The widespread adoption of CI/CD today is heavily influenced by Agile Methods and Extreme Programming methodologies~\cite{beck2000extreme}, which emphasize the automation of software production tasks.
Over the past two decades, CI/CD tools have been widely used to automate development-related activities such as testing, building, quality checking, managing dependencies and detecting security weaknesses~\cite{rostami2023CI,vasilescu2015quality,hilton2016usage,beller2017oops}.

\actions (\gha) was launched in November 2019 by GitHub, the largest collaborative software development platform.
Within 18 months, \gha became the dominant CI/CD solution for GitHub repositories~\cite{Golzadeh2022-CI}.
Decan \etal~\cite{decan2022use} reported that by the end of January 2022, \gha had achieved an adoption rate of 43.9\% in a dataset of 68K GitHub repositories.

Like many other CI/CD tools, \gha requires developers to define workflows in configuration files to automate a repository's CI/CD pipeline.
Following the \emph{Configuration as Code} (CaC) practice, these configuration files are stored in a human-readable \yaml format within the repository.
As such, workflow files are subject to version-controlled changes throughout the repository's lifetime.
These changes are driven by evolving software project requirements, technological advancements, and the shifting needs of developers~\cite{mazrae2023preliminary}.

Previous studies have sought to understand changes in workflow files for different CI/CD tools~\changed{\cite{mazrae2023preliminary,gallaba2018use,durieux2019analysis,zampetti2020empirical,zampetti2021ci,valenzuela2022evolution}}, with an important focus on Travis CI, as it was the dominant CI/CD tool on GitHub before \gha's introduction~\cite{Golzadeh2022-CI}.
We are not aware of any large-scale study specifically focused on the types of syntactic changes made to \gha workflow files.
\changed{The analysis in} this paper aims to close this gap, and \changed{its results} could be used to suggest various improvements to maintaining \gha workflows, such as
following best practices,  increasing the awareness and take-up of existing tools (such as Dependabot for managing workflow dependencies),
improving automated tool support (e.g. for debugging, testing and refactoring workflows). The results can also instruct other researchers to conduct future work in this domain.

\changed{We carry out a} mixed-method empirical analysis, \changed{formulated around} four research questions:

\begin{description}
    \item[RQ1] \emph{How frequently are workflow files changed?}
    We aim to identify and quantify how frequently workflow files are added, modified, renamed, or removed.
    Our results show that, while around 25\% of repositories add more than one workflow file over their lifetime (with a median of three files per repository),
    modifications are the most common type of change to workflow files.
    On average, workflow files are updated every 159 days, with 7.3\% of all files modified each week.
    Renaming and removal are much less common.
    \changedTo{Moreover, most removals tend to occur shortly after the workflows' creation, suggesting the experimental nature of such workflows.
    We also identify bursts (\ie rapid successions) of workflow changes, suggesting the need for improved debugging and testing tools.
    We found no conclusive evidence that the introduction of LLM coding tools or other major technological changes might have impacted the workflow change frequency or burst behaviour.}

    \item[RQ2] \emph{Which conceptual changes are made to workflows?}
    Through a qualitative manual inspection of commits modifying workflow files, we classify changes into seven high-level \emph{concepts} of related syntactic entities.
    Our analysis of 439 commits reveal 1,109 individual changes dominated by modifications inside workflow files (53.4\%).
    Most of these modifications (63.7\%) belong to the concepts of task specification and task configuration, suggesting that maintainers primarily modify workflow steps to refine or extend functionality.
    \changedTo{Commits touching more than one workflow are relatively common but tend to apply similar routine maintenance-related changes to multiple workflows.}

    \item[RQ3] \emph{What types of changes are made to workflows?}
    We quantitatively analyze to which extent workflow entities are being added, removed, or modified during the evolution of workflows.
    Our results show that modifications dominate and this proportion continues to increase over time.
    Nonetheless,  %
    changesets combining additions, removals, and modifications remain relevant for a non-negligible fraction of cases.
    \changedTo{We found no conclusive evidence that the introduction of LLM coding tools or other major technological changes might have impacted the change frequency of fine-grained changes to workflow entities.}

    \item[RQ4] \emph{Which syntactic entities are frequently changed in workflows?}
    Building on the results of RQ2 and RQ3, we quantify the most \changed{frequent} syntactic entities and workflow paths that are subject to changes.
    Our analysis shows that changes concentrate on task specification and task configuration, with the majority occurring in the workflow jobs, and more particularly its steps.
    Renaming fields and creating or updating workflow matrix strategies also play a notable role.
 \end{description}

\noindent In general, this article has three main contributions:
    (i) we provide catalog of seven types of conceptual changes in \gha workflow files based on a manual \emph{qualitative} analysis of 439 workflow file modifications;
   (ii) we report on our finding of a large-scale quantitative analysis over \changed{267K+ workflow change histories from workflow files across 49K+ public GitHub repositories covering 3.4M+ workflow file versions from November 2019 to August 2025};
    (iii)
    \changed{we relate our findings to previous work in the context of CI/CD and GitHub Actions, highlighting observed changes, assessing potential improvements, and suggesting how to further advance research and  practice in the GitHub workflow ecosystem.
}

The remainder of this article is structured as follows:
\sect{sec:related} reviews related work, covering empirical research on CI/CD tools, the evolution of CI/CD configuration files, and \actions.
\sect{sec:methodology} describes the methodology and dataset used for the study.
Sections~\ref{sec:rq1}, \ref{sec:rq2}, \ref{sec:rq3}, and \ref{sec:rq4} present the results of the empirical analysis of changes in \actions workflow files.
\changed{\sect{sec:discussion} discusses the research findings, placing them in the context of related work on CI/CD and GitHub Actions.
}
\sect{sec:threats} outlines potential threats to the validity of \changed{the} study.
Finally, \sect{sec:conclusion} concludes the article and provides directions for future work.
\section{Related Work}
\label{sec:related}

This section reviews the existing literature on the usage of CI/CD in software projects (\sect{subsec:related_CI/CD_general}), the evolution of CI/CD tool usage prior to the advent of \gha (\sect{subsec:related_evolution_cicd}), and the evolution of \gha usage specifically (\sect{subsec:related_GHA}).

\subsection{Empirical research on CI/CD}
\label{subsec:related_CI/CD_general}
Systematic literature reviews (SLRs) provide valuable entry points to CI/CD usage practices.
The SLRs by Shahin \etal~\cite{Shahin2017} and Soares \etal~\cite{soares2022effects} reviewed the scientific literature on the implementation, benefits, challenges, and shortcomings of CI/CD practices across various environments.
These SLRs did not include any research published after 2019, thus excluding \gha, which was introduced in November 2019.

Shahin \etal~\cite{Shahin2017} reviewed 69 scientific articles published up to 2016, synthesizing reported approaches, tools, challenges, and practices for adopting and implementing continuous practices.
The studies highlighted increased adoption of continuous practices, integration problems, and benefits such as reduced build and test time, improved visibility and awareness of results, and enhanced deployment pipelines regarding security, scalability, dependability, and reliability.
Soares \etal~\cite{soares2022effects} examined 101 scientific articles on CI/CD usage published before 2019, identifying empirical evidence on the impact of CI/CD on software development.
They observed a correlation between CI/CD usage and improved productivity, efficiency, and developer confidence.
CI/CD practices promote faster iterations, stability, predictability, and transparency, and benefit pull-based development by improving integration processes.

Several case studies have explored CI/CD usage, cost, and benefits in the context of companies:
Chen~\cite{Chen2015CD, Chen2017CD-JSS} reported on the benefits and challenges of continuous delivery (CD) practices at Paddy Power, including accelerated time to market, improved productivity and efficiency, increased release reliability, and enhanced product quality and customer satisfaction.
Betz \etal~\cite{betz2013implementing} studied the impact of adopting a CI/CD tool in developing AMBER, a molecular dynamics software package.
They reported improved collaboration and communication among globally distributed developers and real-time reporting of failures and benchmark information.
Lu \etal~\cite{lu2014implementation} conducted a case study on D5000, a smart grid scheduling support system, showing that CI and automated testing effectively resolved quality and integration issues without significant overhead.
Kulas \etal~\cite{kulas2014practical} highlighted how CI/CD practices reduced development time for ARGOS,
a software system for processing images produced by a telescope.
Using \jenkins for automated testing ensured the correctness of changes under strict time constraints.
Gmeiner \etal~\cite{gmeiner2015automated} examined CI/CD tool usage in an Austrian online business company, addressing technical and organizational challenges over six years of maintaining an effective CD pipeline.
Savor \etal~\cite{savor2016continuous} studied CI/CD usage at Facebook and OANDA, revealing limitations in fully utilizing continuous deployment due to policy constraints, leading to delays in delivering new features.
Jin~\etal~\cite{10589850} performed a case study of using CI/CD at ByteDance,
observing that the introduction of configuration files improved the reliability and flexibility of CI/CD pipelines.
It encouraged users to build and deploy more frequently and resulted in CI/CD pipelines with fewer steps, higher build frequency, longer build duration, higher success rate, and higher change frequency.
Elazhary \etal~\cite{Elazhary2022TSE} identified benefits and challenges of CI/CD practices in three software development organizations, based on interviews with 18 employees.
Benefits included minimizing merge conflicts, increasing build consistency and reproducibility, and enhancing feedback.
Challenges included difficulties in UI testing, longer build times, PR review bottlenecks, scalability issues, and increased maintenance effort.

Others have studied the impact of CI/CD on open source software (OSS) development.
Zampetti~\etal~\cite{7962383} studied the usage of static analysis tools to enable early detection of potential faults, vulnerabilities, and code smells through their adoption of CI pipelines.
By analysing 20 Java OSS projects hosted on \github and using \travis,
they showed that static analysis tools are used to induce failing builds to
highlight non-adherence to coding standards and missing licenses.
Build failures to highlight potential bugs or vulnerabilities occur less frequently, and in some cases, such tools are activated in a ``softer'' mode, without making the build fail.
The study also reveals that the aforementioned build breakages are quickly fixed by actually solving the problem, rather than by disabling the warning, and are often properly documented.
Hilton~\etal~\cite{hilton2016usage} looked into the usage, costs, and benefits of CI/CD in 34,544 OSS projects on \github.
They observed that CI/CD is widely adopted by the most popular projects, with an increase in the adoption of CI/CD over time.
CI/CD was observed to help projects release more often, and a wide variety of CI/CD tools were used by the projects, with \travis, \circleci, \appveyor, CloudBees, and Werker being the most popular ones.
Vasilescu~\etal~\cite{vasilescu2015quality} studied the quality and productivity outcomes ensuing from embracing CI/CD practices in OSS projects on \github.
They observed that teams using CI/CD are significantly more effective at merging pull requests submitted by core members.
CI/CD was also associated with external contributors having fewer pull requests rejected.
Moreover, core developers in teams using CI/CD tools discover significantly more bugs than in teams not using
such tools.

The mentioned studies are among many that have explored the benefits, challenges, and practices of CI/CD in software projects.
These studies have demonstrated that CI/CD practices can significantly enhance productivity, efficiency, and quality in software development.
However, they have not focused on the evolution of CI/CD configurations over time, which is the focus of the following section.

\subsection{Evolution of CI/CD usage prior to GitHub Actions}
\label{subsec:related_evolution_cicd}

Many CI/CD tools that were in popular use long before the advent of \gha
rely on (often \yaml-based) configuration files for their CI/CD configuration pipelines.
Examples of such tools are \travis, \jenkins, \gitlabci, and \circleci.
Versioning CI/CD configuration files is an instance of the wider practice known as \emph{Configuration as Code} (CaC), which involves encoding configuration settings in a human-readable format (\eg \yaml) and syntax, aligning configuration management with modern software development practices.
CaC provides better opportunities for versioning, code reviewing, and change automation.
In the context of CI/CD, CaC allows configurations to be updated, tested, and deployed automatically alongside application code, ensuring that application and infrastructure changes are coordinated.

The popularity of \travis on \github before the advent of \gha has led many researchers to study this CI/CD tool.
Gallaba and McIntosh~\cite{gallaba2018use} investigated the usage and misuse of features in \travis configuration files in 9,312 \github repositories.
They found that \emph{job processing nodes} were the most frequently modified, indicating that \travis was predominantly used for CI rather than CD. They also developed tools to identify and remove anti-patterns in \travis configuration files.
Similarly, Vassallo \etal~\cite{vassallo2019automated} created a tool to detect anti-patterns in Java projects using \travis.
They based their identification of critical anti-patterns on Duvall's work~\cite{duvall2007continuous}, which served as a benchmark for quality checking in continuous integration.

Durieux \etal~\cite{durieux2019analysis} compiled a dataset of over 35 million \travis jobs from 272,917 projects.
They discovered that the majority of the 709,000+ commits that modified \travis configuration files were related to debugging.
This finding highlighted the need for more in-depth analysis of the nature of these changes.
Zampetti \etal~\cite{zampetti2020empirical} identified 79 bad CI practices through semi-structured interviews with 13 experts and an analysis of over 2,300 Stack Overflow posts.
They also studied the evolution of changes to \travis configuration pipelines, finding that jobs and steps were the most frequently changed components, and noted an increasing adoption of Docker over time~\cite{zampetti2021ci}.

Apart from \travis, many other CI/CD tools have been used in \github repositories.
Golzadeh \etal~\cite{Golzadeh2022-CI} studied the adoption of CI/CD tools in over 91,000 GitHub repositories related to npm packages.
They found that by May 2021, more than 50\% of the repositories used CI/CD tools, with \actions and \travis being the most prevalent.
Remarkably, \actions replaced \travis as the leading CI/CD tool within 18 months of its introduction, and many repositories transitioned from \travis to \actions.

\subsection{Popularity and Usage of GitHub Actions}
\label{subsec:related_GHA}

The introduction of \gha significantly transformed the CI/CD landscape on \github.
The wide array of features and integrations offered by \gha, particularly the concept of Actions as reusable components in CI/CD workflows, proved to be a game-changer.
These Actions can be introduced not only by \github but also by the community, enhancing the flexibility and utility of the platform.

Numerous studies have investigated the usage and evolution of \actions in \github repositories.
Kinsman \etal~\cite{kinsman2021software} analyzed the impact of \gha adoption in 3,190 \github repositories and investigated how developers use Actions and how several activity indicators change after their adoption.
They reported an increase in rejected pull requests and a decrease in commits in merged pull requests.
Their manual inspection of 209 \gha-related issues revealed that developers generally had a positive perception of \gha and the use of Actions.
These findings were corroborated by Chen \etal~\cite{chen2021let} in a replication study involving 6,246 repositories.
Decan \etal~\cite{decan2022use} examined the use of \gha in nearly 70,000 GitHub repositories, finding that 43.9\% used \gha workflows.
They characterized these workflows by their use of jobs, steps, and reusable Actions, demonstrating that most workflows were used to automate the development process rather than deployment automation.
They also found that the majority of reused Actions were provided by \github itself, primarily for checking out the repository or setting up a development environment.
Rostami Mazrae \etal~\cite{rostami2023CI} investigated the reasons behind CI/CD tool adoption, co-usage, and migration in software projects.
Their study highlighted \gha's dominance due to its strong GitHub integration, ease of use, extensive marketplace of Actions, and the generous free tier provided for all users and open-source projects.

Valenzuela-Toledo and Bergel~\cite{valenzuela2022evolution} conducted a preliminary study on the usage and maintenance of \gha workflows in ten popular GitHub repositories, analyzing 222 commits to propose an initial taxonomy of workflow modifications.
Saroar \etal~\cite{saroar2023developers} surveyed 90 producers and users of reusable Action components to understand motivations and best practices in using, developing, and debugging Actions, as well as associated challenges.
They found that users preferred Actions with verified producers and more stars and often switched to alternative Actions when facing issues.
Wessel \etal~\cite{Wessel2023-chapter} suggested studying \gha as a software ecosystem facing challenges similar to traditional software library ecosystems~\cite{decan2019empirical,decan2018impact}.
Moreover, Onsori Delicheh~\etal~\cite{10.1145/3643991.3644899} studied security issues in reusable JavaScript Actions in
\github workflows and showed that more than 54\% of the studied Actions contain at least one security weakness, and a small subset of these weaknesses recur frequently in their code.
For example, 7 out of the top 10 most frequent weakness types are associated with CWE-20 (Improper Input Validation).
They observed that a huge amount of \github repositories are potentially exposed to security issues in their associated workflows.

Rostami Mazrae \etal~\cite{mazrae2023preliminary} examined workflow file changes but did not delve deeply into the specific modifications within the workflow files.
They showed that workflows are subject to the laws of continuing change and continuing growth~\cite{lehman1996laws} and that modification of the contents of workflow files is the dominant type of changes.
Huang \etal~\cite{huang2023cigar} developed a tool that recommends Actions for \gha workflows based on static information of Action usage.
The effectiveness of such tools could be further enhanced by considering the evolution of workflows and the changes they undergo over time.
Zhang \etal~\cite{zhang2024effectiveness} studied the effectiveness of large language models in producing correct workflows, detecting syntactic errors, and identifying code injection vulnerabilities.
Their work could benefit from labeled data categorized by types of changes through time to improve the model performance.
Valenzuela-Toledo \etal~\cite{valenzuela2024hidden} conducted a large-scale empirical investigation to characterize the maintenance of \gha workflows, examining the evolution of workflow files in 183 mature \github projects across ten programming languages.
They found that while \gha improves efficiency, automation introduces hidden costs that need to be properly managed.
As a result, practitioners must plan and allocate sufficient resources for maintaining these workflows, including the identification and documentation of best practices.
Khatami \etal~\cite{khatami2024catching} investigated \gha workflows to identify potential smells.
They reported 22 distinct smells, categorized into three main groups: security, performance/optimization, and general CI/CD smells.
They developed a tool to automatically detect these smells in \gha workflows.
Although their study touched on patterns of frequent changes in \gha workflows, they did not explore these patterns in depth.

In this study, we aim to gain a more comprehensive understanding of the evolution of changes made to \gha workflows by combining qualitative and quantitative analyses of workflow changes over time.\section{Methodology} \label{sec:methodology}

Building further on existing research (\sect{sec:related}), \changed{this article aims}
to understand how \gha workflows change over time, focusing \changed{initially on file-level changes, after which changes to the actual workflow contents are explored}.
A mixed methods research design~\cite{mmresearch2015} is adopted, combining  quantitative and qualitative analyses.
\sect{subsec:gha_syntax} \changed{presents} the \gha syntax to allow the reader to understand the structure and components of workflows, and \sect{subsec:dataset} describes the dataset that is used for this study.

The data and code produced to replicate the analysis are available on Zenodo.~\footnote{\changedTo{\url{https://doi.org/10.5281/zenodo.18414913}}}%

\subsection{GitHub Actions syntax} \label{subsec:gha_syntax}

To enable \gha on a \github repository, one or more \yaml files, each describing a single workflow, should be created and stored in the \emph{.github/workflows} folder.
Each workflow is triggered by some event(s) (\eg when a pull request is submitted, when an issue is created) and performs one or more \emph{jobs} that are composed of one or more \emph{steps}. These steps are either specified in terms of the commands they execute (\entity{run} keyword) or by delegating their implementation to so-called reusable \emph{Actions} (\entity{uses} keyword).
\fig{lst:gha_syntax} provides a real example of a workflow file \changed{to automate dependency checks} in the repository of the Apache NetBeans project.\footnote{\url{https://github.com/apache/netbeans/blob/d19b752/.github/workflows/dependency-checks.yml}}

We start by introducing the terminology that will be used throughout this paper.
A workflow \yaml file is merely a collection of key-value pairs with support for nested mappings and nested sequences. As such, a workflow is a kind of \emph{tree}. We refer to tree nodes as \emph{entities}, and to their specific position in the tree as a \emph{path}, which is a dot-separated sequence of entity names and sequence indexes. For example, we can access the value of the \entity{name} entity of the first step of \fig{lst:gha_syntax} (line 17) through the path \verb|jobs.base-build.steps[0].name|, indicating that the entity can be reached by the sequence composed of the \entity{jobs} entity, the \entity{base-build} entity, the first item\footnote{We start counting from zero, hence \textsf{steps[0]}} in the \entity{steps} entity, and finally the \entity{name} entity.
Most of the keys of workflow entities are imposed by the \gha syntax specification.\footnote{\url{https://docs.github.com/en/actions}}
For those that are not syntactically imposed, such as a job id (\eg \texttt{base-build}) or a parameter name (\eg \texttt{persist-credentials} in \verb|jobs.base-build.steps[0].with.persist-credentials| on line 20), we will occasionally use a \emph{generic} notation such as \verb|jobs.<id>.steps[<nr>].with.<parameter>|.

\begin{figure}[!h]
\footnotesize
\begin{lstlisting}[alsoletter=-,breaklines=true, postbreak=\mbox{$\hookrightarrow$\space},numbers=left,frame=single,basicstyle=\footnotesize,keywords={on,name,workflow_dispatch,concurrency,group,cancel-in-progress,defaults,run,shell,jobs,base-build,runs-on,steps,uses,with,java-version,distribution,persist-credentials,submodules,show-progress,timeout-minutes}]
name: NetBeans Dependency Checks
on:
  workflow_dispatch:
concurrency:
  group: |
    dep-checker-${{ github.head_ref || github.run_id }}-${{ github.base_ref }}
  cancel-in-progress: true
defaults:
  run:
  shell: bash
jobs:
  base-build:
    name: Check Dependencies
    runs-on: ubuntu-latest
    timeout-minutes: 20
    steps:
      - name: Checkout ${{ github.ref }} ( ${{ github.sha }} )
        uses: actions/checkout@v4
        with:
          persist-credentials: false
          submodules: false
          show-progress: false
      - name: Set up JDK
        uses: actions/setup-java@v4
        with:
          java-version: 21
          distribution: "zulu"
      - name: Check Dependencies
        run: |
          DEPS=org.apache.maven:maven-artifact:3.9.9,org.apache.maven.indexer:search-backend-smo:7.1.5
          mvn eu.maveniverse.maven.plugins:toolbox:gav-copy-transitive -Dgav=$DEPS -DsinkSpec="flat(./lib)"
          echo "<pre>" >> $GITHUB_STEP_SUMMARY
          java -cp "lib/*" .github/scripts/BinariesListUpdates.java ./ | tee -a $GITHUB_STEP_SUMMARY
          echo "</pre>" >> $GITHUB_STEP_SUMMARY
          rm -Rf lib
\end{lstlisting}
\caption{Example of a GitHub Actions workflow file \textsf{dependency-checks.yml} taken from the Apache NetBeans project.}
\label{lst:gha_syntax}
\end{figure}

Revisiting the example of \fig{lst:gha_syntax}, the \entity{workflow\_dispatch} entity (line 3) signals that users can manually trigger the workflow to run.
The workflow defines a single job with id \textsf{base-build} (line 12) and name ``Check Dependencies'' (line 13).
It runs on the latest version of Ubuntu (line 14)
and has a timeout of 20 minutes (line 15) to avoid prolonged or stalled runs.
The job is composed of three sequential steps. %
The first step (lines 17-22) checks out the repository's code, using the \textsf{actions/checkout} action (line 18) without storing credentials (line 20) nor cloning submodules (line 21).
The second step (lines 23-27) uses the \textsf{actions/setup-java} action to set up a Java runtime environment for version 21 of the Zulu distribution.
The third step (lines 28-35) performs dependency checks on specific Maven artifacts by running a series of command-line instructions (\entity{run} on lines 29-35).
Finally, the \entity{concurrency} control (lines 4-7) prevents multiple instances of the same workflow from running simultaneously on the same branch, and ensures that ongoing runs are cancelled if a new one starts (line 7).

\subsection{Dataset} \label{subsec:dataset}

To conduct a large-scale empirical analysis of workflow changes over time, we need a large collection of \gha \emph{workflow histories}.
To do so, we rely on the \changed{2025-10-09}\footnote{\changed{\url{https://doi.org/10.5281/zenodo.17301952}}} version of a dataset obtained from public \github repositories that were both popular and active at the time of data collection, meaning repositories with a star count above a certain threshold, a high number of commits, and recent activity~\cite{cardoen2024dataset}.
The dataset contains \changed{267,955 workflow histories obtained from 49,258 \github repositories, accounting for 3,418,911} workflow file snapshots.
It includes all the commits that were made to these workflows, from their introduction in the repositories to their removal (if any) or up to \changed{the data collection date of 25 August 2025}.
Among \changed{others}, the dataset provides, for each workflow history, a unique \textsf{uid} to keep track of renamed workflow files
and, for each commit, its \textsf{date}, the \textsf{name} of the workflow file before and after the commit, and a \textsf{hash} value of the file contents before and after the commit. Since the dataset also contains the contents of all workflow files, this hash value can be used to compare the file contents before and after each commit.
We \changed{apply} two additional filters to this dataset, \changed{motivated by the need to obtain valid workflow histories and ensure consistent temporal coverage}:
\begin{enumerate}
  \item \changed{\textbf{Ensuring valid workflow histories.} 776,339  of the 3,418,911 workflow file snapshots} are flagged as invalid \yaml files. These cases are problematic since \changed{they cannot be parsed} in order to detect the changes made to them.
  Removing only these invalid files does not suffice as it would lead to incomplete workflow histories. Consider for example a workflow history of consecutive workflow files \textsf{A}, \textsf{B} and \textsf{C}, where \textsf{B} is not a valid \yaml file. Removing \textsf{B} from this workflow history would erroneously aggregate the changes from \textsf{A} to \textsf{B} and from \textsf{B} to \textsf{C} into a single set of changes from \textsf{A} to \textsf{C}.
  \changed{To avoid such problems,} we excluded from the dataset the \changed{30,922} workflow histories containing an invalid \yaml file.

  \item \changed{\textbf{Ensuring consistent temporal coverage.}} Since the analysis is based on weekly observations, %
  we  restrict the observation period to complete weeks only, from Sundays to Saturdays.
  We excluded the last \changed{two} days from the dataset so that the observation period covers \changed{316} complete weeks, starting on Sunday 4 August 2019 and ending on  \changed{Saturday 23 August 2025}.

\end{enumerate}

After these steps, the final dataset comprises \changed{236,775 workflow histories from 47,488 repositories, accounting for 2,640,584 workflow file snapshots (abbreviated to workflow files in the remainder).}

\begin{figure}[!h]
    \centering
    \includegraphics[width=\figsize]{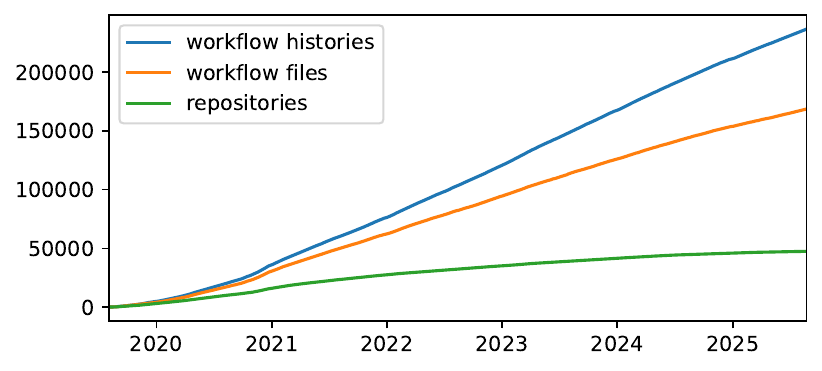}
    \caption{Evolution of the number of \changed{\github} repositories, \changed{workflow files}, and workflow histories in the dataset.}
    \label{fig:number_of_repos_workflows}
\end{figure}

\fig{fig:number_of_repos_workflows} shows the evolution of the number of \changed{\github} repositories, \changed{workflow files} and workflow histories in the dataset.
One can observe that the number of repositories using \gha is increasing through time, and the number of workflow \changed{files} and workflow histories are growing at a faster pace, indicating that more and more workflows are being added to the repositories.
The slight variations that can be observed in late 2020 coincides with %
 restrictions imposed by \travis (a competing CI/CD service) on its free plan, leading many repositories to migrate from \travis to \gha~\cite{Golzadeh2022-CI,mazrae2023preliminary}.

\section{RQ1: How frequently are workflow files changed?} \label{sec:rq1}

The first research question aims to quantify to which extent workflow files are subject to changes during their lifetime. To do so, we consider four different change types, namely \emph{addition} of a new workflow file to the repository, \emph{removal} of the workflow file from the repository, \emph{modification} of the workflow file contents w.r.t. its predecessor in its workflow history, and \emph{renaming} the workflow file w.r.t. the previous one in its workflow history.
For each workflow file in the dataset, we compute its change type w.r.t. its predecessor in its workflow history.

\begin{figure}[!h]
  \centering
  \includegraphics[width=\figsize]{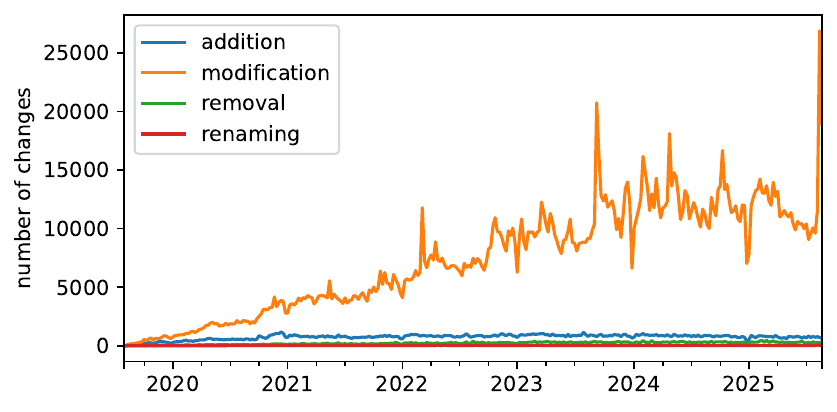}
  \caption{Weekly number of workflow file changes, per change type.}
  \label{fig:number_of_file_level_changes_through_time}
\end{figure}

Overall, we found \changed{2,330,529 modifications, 236,775 additions, 68,083 removals, and 29,159 file renamings}.
\fig{fig:number_of_file_level_changes_through_time} breaks this down into the weekly number of observed workflow file changes.
The overwhelming majority of workflow file changes are modifications, more than a fivefold of all other change type occurrences combined.
The number of modifications tends to increase through time, which is probably a consequence of the fact that more and more workflows are added through time (as observed in \fig{fig:number_of_repos_workflows}).
Additions and removals tend to remain quite stable through time, which is a consequence of the fact that a workflow can \changed{only be added or removed once throughout its history}.
We found
that \changed{24.9}\% of the repositories have more than one workflow file added to them,
with a %
median of \changed{three} workflow files per repository.

Looking more closely at the changes through time in \fig{fig:number_of_file_level_changes_through_time}, we observe some peaks and troughs at specific times, regardless of the change type.
The troughs coincide with the end of year holidays, during which developers are less likely to work and change their workflows.
The peaks will be examined in more detail as part of RQ3 and \changed{RQ4} that delve deeper into the changes to the workflow contents and to the values of workflow entities.

\begin{figure}[!th]
  \centering
  \includegraphics[width=\figsize]{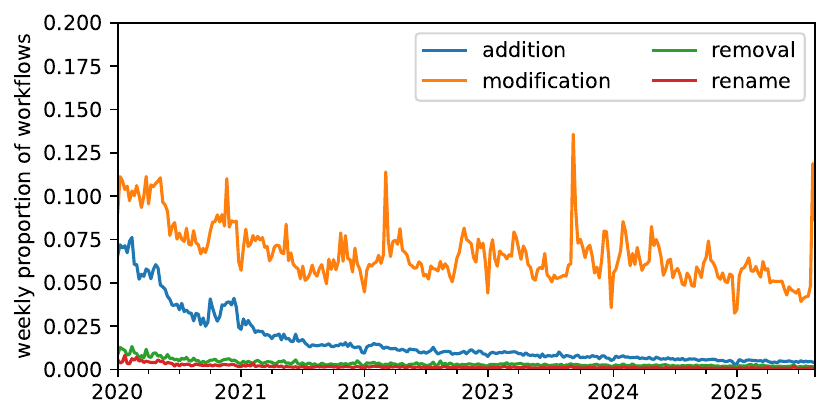}
  \caption{Weekly proportion of workflow files exhibiting a change.}
  \label{fig:proportion_of_workflow_file_changes_through_time}
\end{figure}

To determine whether the observed increase in modifications is mainly due to more workflow files being added over time, or rather to workflow files being modified more frequently, we repeated the previous analysis by computing the weekly proportion of workflow files exhibiting a change of a given type.
\changed{In \fig{fig:proportion_of_workflow_file_changes_through_time}, the ``weekly proportion'' is calculated as the number of workflow files that experienced a specific type of event (addition, modification, removal, or renaming) during a given week, divided by the total number of workflow files that existed in that week. This normalization allows to assess the relative likelihood of workflows undergoing a change of a given type, independent of the overall growth in the number of workflows.
On average, per week 7.3\% of the workflow files are modified, 2.1\% are added, 0.4\% are removed, and 0.2\% are renamed.}

\changed{To answer RQ1 we examined the average change rate of workflow files.
To do so, we identified all commits touching a particular workflow file, and  we divided the workflow's lifespan by the number of times it was touched.
We observed that, on average, workflow files are updated every \changed{159} days,
with 25\% (Q1) of workflow files being touched every \changed{31} days, and 50\% (median) every \changed{82} days.}

\begin{custombox}
\changed{Around 25\% of all repositories add more than one workflow file throughout their lifetime, with a median of three workflow files per repository.
Workflow files are updated on average every 159 days, with 7.3\% of workflow files being modified each week.
While workflow files can be added, modified, renamed or removed, modifications clearly dominate, whereas renaming and removal are uncommon.
}
\end{custombox}

\begin{figure}[!t]
  \centering
  \includegraphics[width=\figsize]{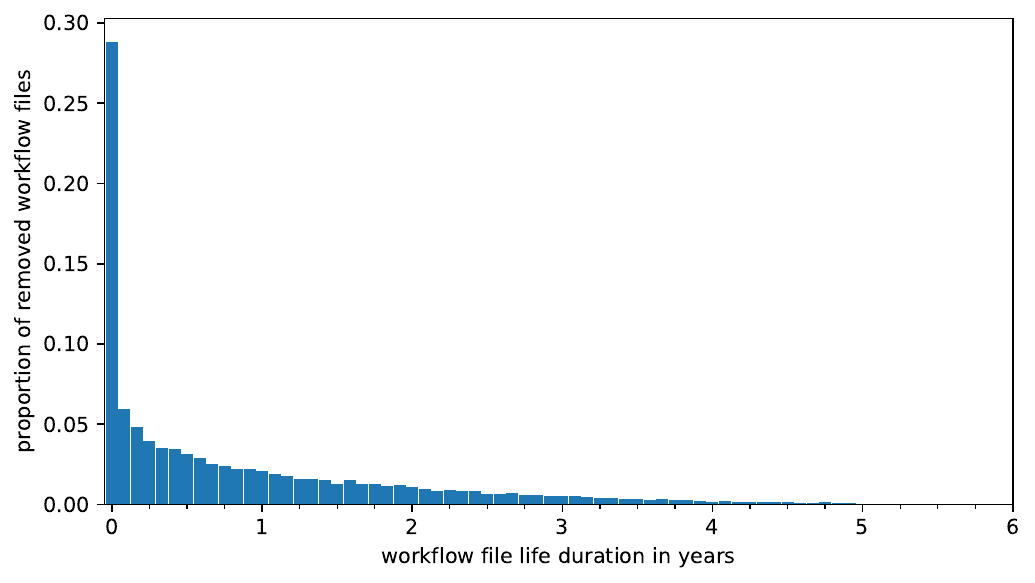}
  \caption{\changed{Proportion of removed workflow files with respect to their lifespan.}}
  \label{fig:proportion_of_removed_workflows_wrt_their_life_duration}
\end{figure}

\changed{\paragraph{Short-lived workflow files}

While the overall frequency of workflow file removals is low, examining when these removals occur provides additional insight into workflow maintenance behavior.
In particular, we aimed to understand whether workflow files that are removed tend to be short-lived or long-standing configurations files.
To this end, \fig{fig:proportion_of_removed_workflows_wrt_their_life_duration} shows the distribution of removed workflow files relative to their lifespan (\ie the duration between the commit that introduced the workflow file and the commit that removed it).
We grouped these durations into monthly intervals and computed their relative proportions (i.e., normalized by the total number of removed workflow files).
The figure allows to examine not just the absolute number of removals, but their relative prevalence across different lifespan ranges.
One can observe that a substantial fraction of workflow files are removed shortly after their creation. Specifically, 15.2\% of all removed workflow files were deleted within the first day, 20.6\% within the first week, and 29.0\% within the first month.
In total, 68,083 workflow files were removed at some point in their lifetime, a significant fraction of which were removed soon after their creation.
Given the short lifespan and limited impact of short-lived workflows on long-term workflow maintenance efforts, we recommend that future longitudinal studies aiming to characterize sustained maintenance practices consider excluding such ephemeral workflow files.
However, we also recognize the value of studying short-lived workflows in their own right, for example, to understand exploration, onboarding patterns, or rapid prototyping activities on \github software projects and their associated workflows.
}

\begin{custombox}
\changed{Removing workflow files is uncommon, and such removals tend to occur shortly after the workflows' creation.}
\end{custombox}

\begin{table*}[!t]
  {
  \footnotesize
  \begin{center}
  \begin{threeparttable}
  \caption{\changed{Workflow burst statistics under varying commit time intervals, showing prevalence, burst count, and commit density.}}
  \label{tab:burst_thresholds}
  \begin{tabular}{r|r|r|r|r}
  \textbf{} & \changed{\textbf{\# workflow}}  &  \changed{\textbf{\% workflow}}&\changed{\textbf{mean}} & \changed{\textbf{mean}} \\
  \textbf{time} & \changed{\textbf{histories}}  &  \changed{\textbf{histories}} &\changed{\textbf{\#bursts}} & \changed{\textbf{\#commits}} \\
  \textbf{interval} & \textbf{with bursts} & \textbf{with bursts} & \textbf{per workflow} & \textbf{in bursts}  \\
  \toprule
    15 mins & \changed{20,277} & \changed{15.37}\% & \changed{1.52} & \changed{3.90} \\
    30 mins & \changed{25,488} & \changed{19.31}\% & \changed{1.57} & \changed{4.05} \\
    60 mins & \changed{29,932} & \changed{22.68}\% & \changed{1.59} & \changed{4.12} \\
   \end{tabular}
    \end{threeparttable}
  \end{center}
  }
\end{table*}

\changed{\paragraph{Commit bursts}
While we have showed that workflow files are frequently modified throughout their lifetime, these aggregate statistics do not reveal \emph{how} such changes are distributed over time.
To better understand whether workflow maintenance occurs as steady, continuous activity or rather as short, intense editing sessions, we examined the temporal clustering of commits touching the same workflow file.

Nagappan \etal~\cite{nagappan2010change} studied commit ``bursts'' in software components as defect predictors.
We investigate whether similar short bursts of rapid commits also occur for workflow files, signaling trial-and-error debugging attempts to fix failing workflows.
To do so, we apply a temporal clustering algorithm on the dataset to identify workflow burst episodes.
We define a burst as a chain of at least three (not necessarily consecutive)
commits belonging to the same workflow history, in which each commit occurs within a fixed short time interval of the previous one. Table \ref{tab:burst_thresholds} reports the results using three thresholds for this time interval: 15, 30, and 60 minutes.
For the 60-minute interval  we observed bursts in 22.68\% (29,932) of all workflow histories, with a relative commit density of  4.13 commits on average per burst (median of 3).
For the shortest interval of 15 minutes, there were still 15.37\% (20,277) of all workflow histories with burst activity.
Across all three time intervals, the mean number of bursts per workflow history remained around 1.55, with each burst typically comprising \changed{3 (50$^{th}$ percentile) to 4 commits (75$^{th}$ percentile).}
We also observed a non-negligible number of longer and more intense bursts. For example, for the 15, 30, and 60 minute intervals, we identified respectively 1,273, 2,120, and 2,818 bursts containing at least 8 commits.
Some more extreme cases involve up to 43 commits~\footnote{First commit of the burst: \url{https://github.com/FreeTubeApp/FreeTube/commit/1c9959ccfc020475caf1e632c30149cb30cd9b78}} for the 15-minute interval and up to 50 commits~\footnote{First commit of the burst: \url{https://github.com/iptv-org/iptv/commit/21c58cccae1a4c570e40e271cdd88054741bdd57}} for the 60-minute interval.
These findings confirm that bursty, high-frequency workflow commits are not uncommon. %
A substantial number of workflow histories, up to one in ten, show signs of clustered change sequences that may reflect iterative testing and debugging.

Taken together, this temporal analysis supports our hypothesis that workflows are frequently changed through short, iterative editing sessions, likely to fix bugs in response to failing workflows.
Further analysis of workflow execution logs would be necessary to fully validate this trial-and-error behavior, \changedTo{but is considered out of scope of the current paper.}
}

\begin{custombox}
\changed{Bursts of rapid successions of workflow changes are not uncommon, \changedTo{suggesting} the need for improved debugging and testing tools.}
\end{custombox}

\begin{figure}[!t]
  \centering
   \includegraphics[width=\textwidth]{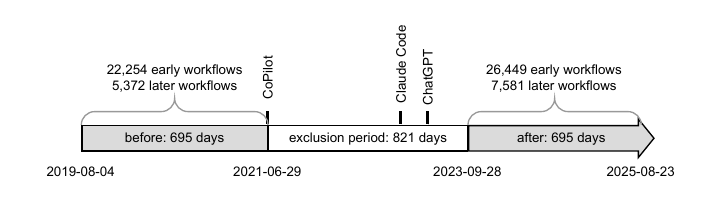}
  \caption{\changedTo{Comparison of early and later-phase workflow histories before and after introduction of LLM-powered coding tools and other major technological changes to  GitHub Actions.}}
  \label{fig:before-after}
\end{figure}

\changedTo{
LLM-powered coding tools such as Copilot, Claude Code and ChatGPT have emerged since 2021 and are becoming increasingly integrated in software development, leading to important shifts in developer roles~\cite{Bird2023-Copilot}.
As a consequence, the emergence and increasing use of such tools might have lead to important changes in how new workflows are created or how existing workflows continue to be maintained over time.
To verify this hypothesis, we carry out a time-based comparison of both kinds of evolution scenarios before and after the emergence of such tools.
As shown in \fig{fig:before-after}, we compare two distinct 695-day periods of our dataset, corresponding respectively to the initial situation of \gha workflow usage \textbf{before} the existence of LLM coding tools, and the more recent situation of workflow usage \textbf{after} such tools have become integrated and used in GitHub repositories. These before and after periods are separated by an exclusion period of 821 days, starting at the day that  Copilot became part of the GitHub platform, and including the emergence of  other AI coding tools and many other technological changes.
Within both periods we select all \textbf{early}-phase workflow histories reflecting the first three months of file changes for a given workflow history since its creation (\ie since the first commit that created the workflow file). %
We also select all \textbf{later}-phase workflow histories reflecting the three months of file changes from month nine to twelve of the workflow history, considering only those workflows with at least one year of activity within the considered period.
We compare three change metrics for both \emph{early}- and \emph{later}-phase workflow histories between the \emph{before} and \emph{after} periods:
\begin{description}
\item[change frequency] is the number of commits touching the workflow during the considered three-month history 
\item[bursts per workflow] counts the number of bursts during the three-month workflow history, focusing only on the subsets of early-phase and later-phase histories that contain at least one burst
 \item[commits per burst] counts the number of commits observed in each burst considered in the previous metric
 \end{description}
As for \tab{tab:burst_thresholds}, we compute three variants of the last two metrics, for burst time intervals of 15, 30 and 60 minutes, respectively.
We apply non-parametric Mann-Whitney U tests to find statistical evidence of differences for (each variant of) each metric, using a significance level of $\alpha=0.01$ after Bonferroni correction to control the family-wise error rate induced by multiple tests applied to the same populations~\cite{halperin1988some}. In those cases where the null hypothesis ($H_0$) can be rejected, we use Cliff's $\delta$ effect size measure  to quantify the magnitude of observed differences, and interpret its value according to Romano \etal \cite{Romano2006}.

For the \textbf{change frequency} metric, $H_0$ is rejected for both early- and later-phase workflows, reflecting a decreased change frequency in the \emph{after} period. However, the effect size is \emph{negligible} ($\delta < 0.147$).
For the \textbf{bursts per workflow} metric, $H_0$ is only rejected when comparing 15-minute burst intervals for early-stage workflows, but the effect size remains \emph{negligible}.
For the \textbf{commits per burst} metric, $H_0$ is never rejected.
Taken together, this comparative analysis does not reveal any conclusive impact of the use of LLM coding tools, or any other important technological change during the exclusion period.
}

\begin{custombox}
  \changedTo{There is no conclusive evidence that the introduction of LLM coding tools or other major technological changes have impacted the workflow change frequency or burst behaviour.}
\end{custombox}

\section{RQ2: Which conceptual changes are made to workflows?} \label{sec:rq2}

RQ1 revealed that a significant proportion of workflow files undergo frequent modifications throughout their lifetime.
With RQ2 we aim to qualitatively analyse the nature and location of the \changed{changes} made to the workflow contents stored in these files. This will allow to assess whether specific kinds of changes are more frequent than others, thus providing the basis for a large-scale quantitative analysis in the next RQs.

We conducted a manual qualitative investigation of the \changed{changes} made to the contents of a \changed{small yet statistically representative} sample of workflow files. \changed{All four authors were actively involved in this process, with the first author leading the design, coordination, and execution of the main tasks.}
We followed the \emph{framework method}~\cite{spencer2004quality} to identify and categorize \changed{conceptual changes to \gha workflows}.

We began by \emph{familiarizing}~\cite{spencer2004quality} ourselves with the syntax and structure of workflows to understand the possible \changed{changes that can be made to them. During this stage of \emph{framework identification}~\cite{spencer2004quality}, we realized that most workflow file modifications could be mapped to changes of specific syntactic \emph{entities} (such as triggers, environment variables, matrix strategies, jobs, and steps).}

\changed{During a subsequent \emph{indexing} phase~\cite{spencer2004quality},} we qualitatively examined a statistically representative number of modified workflow files. We aimed for a 95\% confidence level with a 5\% margin of error~\cite{burmeister2012sample}. This required randomly sampling 389 distinct workflow histories from our dataset (Section~\ref{subsec:dataset}). Each workflow history was chosen from a different repository to ensure diversity. From each workflow history, we arbitrarily selected one workflow file for manual analysis. \changed{Its changes relative to its predecessor were obtained from the git commit information.}

The sample of 389 \changed{cases} was divided into three batches, each randomly assigned to two authors for independent classification. Each author \changed{identified all observed changes to the contents of the assigned workflow file, as well as their change type (\ie addition, modification, or removal), providing clarifying comments when necessary.} At the end of this process, the two authors compared and merged their lists of identified changes and resolved disagreements. \changed{In the few cases where consensus could not be reached, a third author intervened to come to an agreement.
At the end of this step, for each workflow file, we have set of changes like `change an instruction in \entity{run:} step' or `add job permission'.}

\changed{To ensure saturation, we continued the above process by iteratively analysing ten additional workflow file modifications at a time. After processing 50 more cases in this way, we observed that the last batch did not yield noteworthy new information, concluding that saturation was reached. Considering all 439 analysed cases in total (389 + 50), we reduced the margin of error to 4.67\% while keeping the 95\% confidence level.}

\changed{We then \emph{charted}~\cite{spencer2004quality} the documented changes by systematically mapping all changed workflow \emph{entities} into a provisional classification scheme of change types (addition, modification, removal).
For instance, we mapped the observed workflow modification `change an instruction in \entity{run:} step' to a `modification' at `step' level for a `command'.
Similarly, we mapped the observed workflow modification `add job permission' to an `addition' at `job' level for `permissions'.
This charting allowed us to refine and consolidate the initial categories that had emerged during the indexing phase.}

The final step consisted of \emph{interpretation}~\cite{spencer2004quality}, \changed{\ie abstracting the concrete change types towards higher-level conceptual categories.} We employed an open card-sorting approach, allowing broader concepts to emerge from the classification of the \changed{charting phase}, without imposing any pre-defined concepts~\cite{spencer2009card}.
The coding during this step was performed by the first author, followed by a review and discussion with the other authors until a negotiated agreement was reached~\cite{garrison2006revisiting}.

\begin{table*}[!t]
  {\footnotesize
  \begin{center}
  \caption{\changed{Conceptual changes being made to workflows}, together with the number and proportion of occurrences observed in the considered sample of 439 \changed{modified workflow files}.}
  \label{tab:change_catalog}
  \begin{tabular}{l|rr|rrr}
  & \multicolumn{2}{r|}{\textbf{all changes}} &\multicolumn{3}{c}{\textbf{change type}} \\
  \textbf{concept} & \# & \% & \textbf{modifs.} & \textbf{additions} & \textbf{removals} \\

  \toprule
    Task Specification (TS) & 441 & 39.8\% & 196 & 145 & 100 \\
    Task Configuration (TC) & 265 & 23.9\% & 226 & 21 & 18 \\
    Documentation (D) & 201 & 18.1\% & 64 & 87 & 50 \\
    Runtime Configuration (RC) & 85 & 7.6\% & 50 & 26 & 9 \\
    Execution Triggers (ET) & 46 & 4.2\% & 23 & 19 & 4 \\
    Execution Rules (ER) & 43 & 3.9\% & 14 & 16 & 13 \\
    \changed{Formatting (F)} & 28 & 2.5\% & 19 & 2 & 7 \\
    \hline
    \textbf{total} & 1,109 & 100\% & 53.4\% & 28.5\% & 18.1\% \\
  \end{tabular}
  \end{center}
  }
\end{table*}

\changed{The outcome of this step was a set of seven concepts of identified workflow changes}. %
\tab{tab:change_catalog} reports on each of these concepts, providing the number of changes overall and per change type (\ie modification, addition, and removal) in the considered sample of 439 \changed{cases}.
In total, 1,109 change types were observed, each associated with one specific concept.
53.4\% of these changes were \emph{modifications}, 28.5\% were \emph{additions}, and 18.1\% were \emph{removals}.
\changed{Below we discuss each of the seven concepts in detail.}
The first three concepts constitute the large majority (81.8\%) of all observed changes:

\emph{Task Specification (TS)} is the concept referring to the structure and logic that defines what tasks are performed in the workflow, and how they are executed.
Entities belonging to this concept are changed the most frequently (39.8\% of all observed changes). They also correspond to the highest number of additions and removals.
\changed{Changes belonging to this concept include the introduction, removal and reordering of jobs and steps, as well as changing the \entity{uses} and \entity{run} specifications of steps.}
Another frequent category of changes is adding or modifying a matrix strategy (\entity{matrix}, \entity{strategy}, \entity{include} and \entity{exclude} keywords) to allow jobs to run multiple times based on the combination of variables used in the matrix definition.
Moreover, changes of the command of the steps for \entity{run} were also observed frequently.

\emph{Task Configuration (TC)} \changed{is} the concept covering the detailed settings of how individual steps are configured and executed within a workflow job.
Entities belonging to this concept are changed quite frequently (23.9\% of all observed changes) and correspond to the highest number of modifications observed during the qualitative analysis.
\changed{Most of these modifications relate to the configuration of reusable Actions such as changing their versions (\entity{uses} keyword) and their parameters.
Other entities that belong to this concept are}
defining working directories (\entity{working-directory}),
declaring and using jobs and step identifiers (\entity{id}),
and managing outputs of steps to be used by other jobs (\entity{outputs}).

\emph{Documentation (D)} is the concept referring to the practice of \changed{providing} descriptive information that explains or clarifies various parts of the workflow.
Such documentation increases the readability and understandability of the purpose, structure, and behavior of the workflow.
This can be achieved by providing human-readable names (\entity{name}) for workflows, jobs, or steps, or by including comments %
 in the \yaml file.
Entities belonging to this concept are changed quite frequently (18.1\% of all observed changes).
Most of the changes are related to adding or modifying names in different parts of the workflow file.

\bigskip
The next four concepts are considerably less subject to changes, accounting for only 18.2\% of all observed changes.

\emph{Runtime Configuration (RC)} refers to the settings that control \emph{how} and \emph{where} the workflow needs to run.
Proper runtime configuration ensures that workflows are efficient, secure, and adaptable to different project needs or environments.
This includes defining the environment in which jobs will operate, ensuring it behaves as intended when it \changed{is} triggered (\entity{env}), setting global defaults for jobs (\entity{defaults}), specifying the operating system or platform (\entity{runs-on}), and managing the security permissions required for the workflow to perform specific actions (\entity{permissions}).
These settings are crucial for ensuring consistency across environments, and controlling the access the workflow has during execution.
Despite their importance, entities belonging to this concept are changed less frequently (7.6\% of all observed changes).

\emph{Execution Triggers (ET)} refer to the mechanisms that determine \emph{when} a workflow is activated and executed.
These triggers define the specific events or conditions that prompt the workflow to run (\entity{on}), allowing for automated responses to various activities within a repository.
By configuring execution triggers, developers can specify the types of events, such as code pushes (\entity{push}), pull requests (\entity{pull\_request}), or issue comments, that will initiate the workflow.
Additionally, these configurations may include details about the data associated with these events, enabling the workflow to respond appropriately based on the context of the trigger.
Understanding and managing execution triggers is essential for creating efficient and responsive workflows that align with the development lifecycle and project requirements.
However, once declared in the workflow, these entities are changed infrequently (4.2\% of all observed changes), as they are typically set up at the beginning of the workflow development process and remain stable throughout the workflow's lifecycle.

\emph{Execution Rules (ER)} establish the conditions under which workflow jobs are executed, ensuring that jobs behave predictably and efficiently.
They enable workflow maintainers to control the flow of the workflow by specifying prerequisites that must be met before a job runs (\entity{if}), determining the sequence in which jobs are executed (\entity{needs}), and managing error handling strategies (\entity{continue-on-error}).
Execution rules also allow for time constraints on jobs or steps, which can prevent indefinite runs and ensure that resources are managed effectively (\entity{timeout-minutes}).
Only 3.9\% of all observed changes were related to entities in this concept.
The low frequency of changes can be attributed to the reduced need for such conditions in simpler scenarios.

\changed{\emph{Formatting (F)} corresponds to cosmetic changes that do not affect a workflow's execution. }
This includes \changed{cases like} changing the list representation of a workflow trigger, changes to delimiters, and quotation styles (\eg replacing single by double quotes).
\changed{Changes in formatting} were observed very infrequently (2.5\% of all observed changes).

\begin{custombox}
\changed{
1,109 individual workflow changes across 439 commits were manually classified into 3 change types and 7 concept categories.
The most frequent change type was modification of workflow entities (53.4\%), followed by addition (28.5\%) and removal (18.1\%). The large majority of workflow changes (81.8\%) fall into three concept categories: task specification (39.8\%), task configuration (23.9\%), and documentation (18.1\%).
The first two concepts suggest that maintainers mainly change workflow configurations to refine or extend its functionality.
Less frequent changes involve runtime configuration (7.6\%), execution triggers (4.2\%), execution rules (3.9\%), and formatting (2.5\%), which tend to be defined early in a workflow’s lifecycle and remain stable over time.
}
\end{custombox}

\changedTo{

The previous analysis focused on changes to individual workflow files.
However, some GitHub repositories contain multiple workflow files that may be changed within the same commit.
In our sample of 439 commits, 95 of them (21.64\%) touched multiple workflow files.
We performed an in-depth analysis of those 95 multi-workflow commits to identify the nature of their changes.
We found 23 cases of \emph{unrelated changes} where no relation could be discerned between the changes to each workflow,
68 commits with \emph{co-changes} where the same kind of change was made to each workflow,
and 4 commits with \emph{dependent changes} where a change to some workflow was the natural consequence of a different change to another. The commits with \emph{dependent changes} involved moving a job or steps from one workflow to another, splitting a workflow into two, and refactoring jobs into a reusable workflow.
31 of the 68 commits with \emph{co-changes} involved updating the version of reused components, either manually (13 commits)
or automatically (18 commits) via tools such as Dependabot or Renovate.
The 37 remaining \emph{co-change} commits applied the same change to multiple workflow files (changing branch name, updating \entity{run} or updating the matrix strategy).
}
\begin{custombox}
\changedTo{
Most of the multi-workflow changes (68 out of 95) apply similar changes to multiple workflows.
As a consequence, the quantitative analysis in the remaining RQs will focus on changes to individual workflows only.
}
\end{custombox}

\section{RQ3: What types of changes are made to workflows?}\label{sec:rq3}

\changed{RQ2 reported on a manual \textbf{qualitative} classification of} workflow changes into seven different concepts. Each concept encompasses multiple syntactic entities, corresponding to specific workflow keys. %
\changed{RQ3 shifts the focus toward a \textbf{quantitative} perspective aimed at} analysing the number and type of changes made inside workflows (\ie addition, modification and removal), whereas RQ4 will focus on the frequency of changing  specific syntactic entities.
In RQ3, we are particularly interested in understanding whether developers tend to create small changes (e.g., modifying a single step) or larger ones (e.g., modifying multiple steps, adding new steps, etc.).
This distinction provides insights in how developers approach workflow maintenance, particularly whether their changes are focused on small refinements or involve more complex updates that may require significant restructuring.

Answering both RQ3 and RQ4 requires a tool to \changed{compute the changeset of syntactic differences} between a modified workflow file and its immediate predecessor. %
To do so, we rely on \gawd (GitHub Actions Workflow Differ), a Python-based tool that identifies  changes made to workflow files~\cite{mazrae2024gawd}. %
\gawd computes the complete set of changes that can be observed between two workflow files provided as input.
The tool detects \emph{addition} and \emph{removal} of workflow entities, \emph{modification} of the value of an entity, \emph{moving} the position of a step in jobs, and \emph{renaming} jobs (\ie when the \textsf{job} key is changed).
For each of these change types, \gawd provides the corresponding paths and values from the two workflow files.
An example of changeset provided by \gawd for a workflow file and its immediate predecessor is shown in \fig{fig:tool_result_example}.

\begin{figure}[!h]
\centering
\begin{lstlisting}[breaklines=true, postbreak=\mbox{$\hookrightarrow$\space},frame=single,numbers=left,basicstyle=\footnotesize\ttfamily]
renamed jobs.main to jobs.main-task
added env with {'REGISTRY': 'ghcr.io', 'IMAGE_NAME': '${{ github.repository }}'}
moved jobs.main.steps[5] to jobs.main-task.steps[6]
changed jobs.main.steps[5].with.push from "${{ github.repository == 'cloud-hypervisor/rust-hypervisor-firmware' && github.event_name == 'push' }}" to "${{ github.event_name == 'push' }}"
changed jobs.main.steps[5].with.tags from 'rusthypervisorfirmware/dev:latest' to '${{ steps.meta.outputs.tags }}'
removed jobs.main.steps[4].if with "${{ github.repository == 'cloud-hypervisor/rust-hypervisor-firmware' && github.event_name == 'push' }}"
added jobs.main-task.steps[4].with.registry with '${{ env.REGISTRY }}'
removed jobs.main.steps[6] with {'name': 'Image digest', 'run': 'echo ${{ steps.docker_build.outputs.digest }}'}
added jobs.main-task.steps[5] with {'name': 'Extract metadata (tags, labels) for Docker', 'id': 'meta', 'uses': 'docker/metadata-action@v4', 'with': {'images': '${{ env.REGISTRY }}/${{ env.IMAGE_NAME }}', 'flavor': 'latest=true\n'}}
\end{lstlisting}
\caption{Example of \changed{a changeset of syntactic differences} computed by \gawd on a modified workflow file and its predecessor, taken from the popular repository \url{cloud-hypervisor/rust-hypervisor-firmware}.}
\label{fig:tool_result_example}
\end{figure}

We executed \gawd on all modified workflow file snapshots and their immediate predecessors in our dataset.
This resulted in \changed{2,261,806} changesets containing a total of \changed{7,811,892} changes.
The number of changesets is lower than the number of modified workflow files (\changed{2,330,529} cases, see RQ1), \changed{since} \gawd excludes changes that are irrelevant for our analysis, such as adding or removing whitespaces, comments, etc.
The dataset of changes and changesets created in this process is available %
through our replication package.

The \changed{7,811,892} changes can be broken down into \changed{3,551,083} \textsf{modification}s, \changed{1,513,588} \textsf{addition}s, \changed{887,961} \textsf{removal}s, \changed{1,811,068} \textsf{move}s, and \changed{48,192} \textsf{rename}s.
We \changed{ignore} the \textsf{move} and \textsf{rename} change types because they are of little relevance for our analysis and because \gawd only provides a partial view of them. For instance, \gawd only detects \textsf{rename}s of jobs while \textsf{move}s are only detected inside the list of steps.

\begin{figure}[!t]
  \centering
  \includegraphics[width=0.8\textwidth]{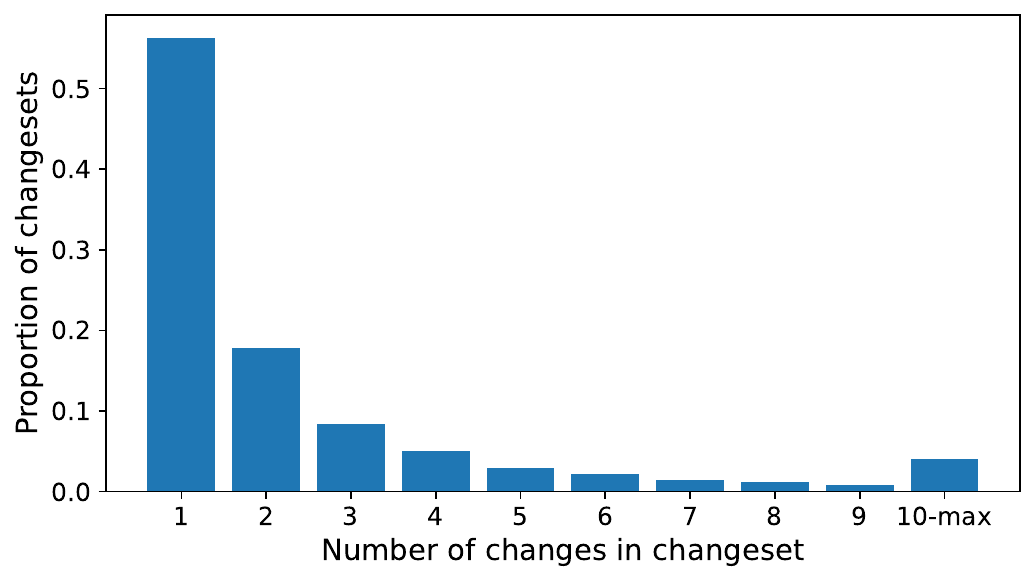}
  \caption{Histogram of the proportion of changesets in function of number of changes.}
   \label{fig:histogram_of_changesets_wrt_number_of_changes_plot}
\end{figure}

The number of changes (\changed{5,952,632} after \changed{ignoring} \textsf{rename}s and \textsf{move}s) is \changed{2.67} times \changed{higher} than the number of changesets (\changed{2,226,452}) because many changesets contain more than a single change. Still, the median number of changes per changeset is one, \changed{implying} that at least half of all changesets involve only a single change.
\fig{fig:histogram_of_changesets_wrt_number_of_changes_plot} \changed{visualises} the distribution of changes per changeset.
We observe that most changesets are small and focused, with a steep decline in the frequency of larger changesets. %
We also observe a group of changesets that contain disproportionately many changes.
Indeed, 10\% of changesets include more than \changed{five} changes, and \changed{nearly 4\%} %
contain ten or more changes,
with a maximum of 1,596 changes observed in some changeset.
Such outliers are typically the result of workflow files being modified \changed{by} automated tools.~\footnote{Example: \url{https://tinyurl.com/3xcjtude}}

Further looking at the change types in these changesets,
we found that \changed{78.98}\% of all the changesets contain only one change type,
with \textsf{modifications} being the most common (\changed{78.81}\%), followed by \textsf{additions} (\changed{15.79}\%) and \textsf{removals} (\changed{5.40}\%).
The remaining \changed{21.02}\% of changesets involve multiple change types. The most frequent combination of change types is \textsf{addition} and \textsf{modification} (\changed{37.43\%} of multi-type changesets).
Other common combinations include all three \changed{change} types together (\changed{28.40}\%), \textsf{addition} and \textsf{removal} (\changed{20.10}\%), and \textsf{modification} and \textsf{removal} (\changed{14.05}\%).

\begin{figure}[!t]
  \centering
  \includegraphics[width=\figsize]{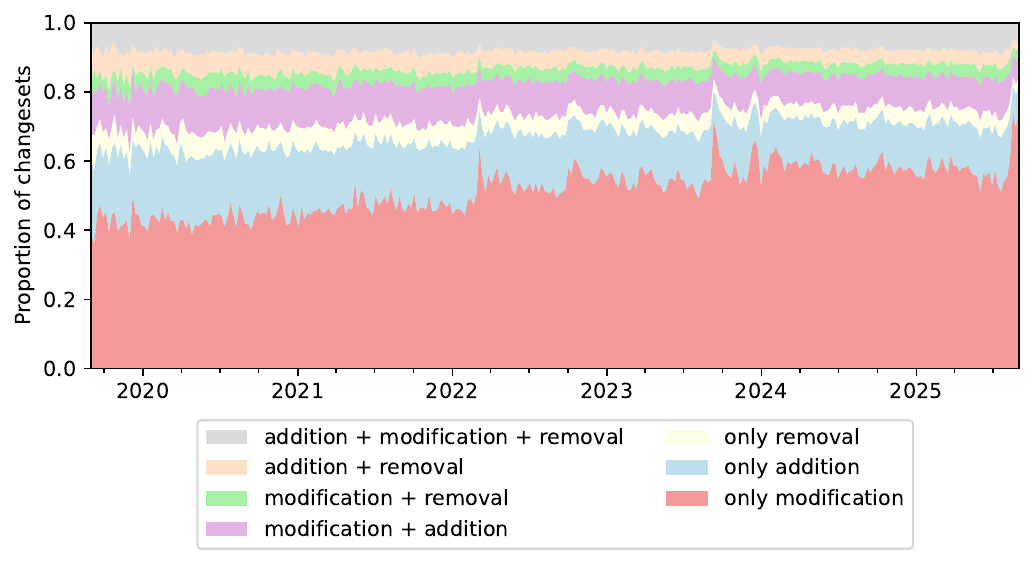}
  \caption{Weekly evolution of the proportion of changesets in function of the change types.}
  \label{fig:percantage_of_events_through_time_area_plot}
\end{figure}

To analyze how these trends evolve over time, we examined the proportion of changesets containing at least one instance of each change type on a weekly basis.
\fig{fig:percantage_of_events_through_time_area_plot} %
reveals a growing dominance of changesets consisting solely of \textsf{modifications}, indicating that as workflows mature, developers focus more on refining configurations than on \changed{adding} or removing elements.
At the end of the observation period, \textsf{modifications} account for \changed{51.29}\% of monthly changesets on average, compared to \changed{15.6}\% for \textsf{additions} and \changed{4.89}\% for \textsf{removals}.
Multi-type changesets continue to remain relevant \changed{as well.}
\changed{This suggests} that, while most changes are focused on refining workflows, a significant portion still involves more complex \changed{changes} combining multiple types, reflecting an ongoing balance between workflow maintenance activities and functional enhancements.

\changed{We also} observe two relative \changed{spikes} in modification activity, one in the first quarter of 2022 and another in the third quarter of 2023,
\changed{aligned} with major events in the GitHub Actions ecosystem.
The 2022 \changed{spike} corresponds to the migration from \textsf{node12} to \textsf{node16}, coinciding with the release of version 3 of several widely used Actions, including checkout\footnote{\url{https://github.com/actions/checkout/releases/tag/v3.0.0}}, setup-node\footnote{\url{https://github.com/actions/setup-node/releases/tag/v3.0.0}}, setup-python\footnote{\url{https://github.com/actions/setup-python/releases/tag/v3.0.0}}, and cache.\footnote{\url{https://github.com/actions/cache/releases/tag/v3.0.0}}
These Actions are among the most frequently used in workflows, as previously reported by Decan et al.~\cite{decan2022use}.
Similarly, the \changed{spike} in 2023 can be attributed to planned deprecations, such as the phase-out of Ubuntu 18.04 and the removal of Python 2.x support in version 3 of setup-python, \changed{which} prompted widespread updates to existing workflows.

\begin{custombox}
  \changed{We analyzed \changed{2,261,806} workflow file changesets, comprising a total of \changed{7,811,892} individual changes obtained through \gawd,} composed of additions, removals and modifications \changed{of workflow entities}.
  We observed that \changed{nearly four out of five} changesets (\changed{78.98}\%) \changed{belong to} a single change type, and in \changed{nearly four out of five} cases (\changed{78.81}\%) these are modifications.
The proportion of changesets consisting exclusively of modifications \changed{tends to grow over time, suggesting an increased focus on refining existing workflows.}
 \end{custombox}
\changedTo{
To assess the possible impact of LLM coding tools on the changes made to the \emph{contents} of workflows, we apply the same analytical procedure as in RQ1.
Specifically, we compare the \textbf{change frequency} of \emph{early}-phase and \emph{later}-phase workflow histories between the \emph{before} and \emph{after} period of our dataset (see \fig{fig:before-after}), measuring change frequency for the five change types to workflow entities: modification, addition, removal, move, and renaming.
$H_0$ is only rejected for the \emph{modification} change type, for both \emph{early}-phase and \emph{later}-phase workflow histories, and for the \emph{addition} change type only for \emph{early}-phase but the effect sizes are \emph{negligible} ($\delta < 0.147$) reflecting a decreased change frequency in the after period.
As a consequence, this fine-grained comparative analysis at the level of change types to workflow entities again does not reveal any conclusive impact of the use of LLM coding tools.}

\begin{custombox}
  \changedTo{There is little to no evidence that the introduction of LLM coding tools or other major technological changes have impacted the frequency of fine-grained changes to workflow entities.}
\end{custombox}

\section{RQ4: Which syntactic entities are frequently changed in workflows?} \label{sec:rq4}

\changed{
Continuing the quantitative analysis of RQ3, RQ4 takes a finer-grained perspective by focusing on the syntactic changes made to workflows.
By doing so, we aim to uncover which workflow entities are most frequently changed.
Understanding these syntactic hotspots can guide future improvements to the workflow language and its supporting tools.
Frequently changed entities are prime candidates for enhanced tool support, such as automated bug detection, refactoring, and security audits, while infrequently changed entities may pinpoint low usage, limited usefulness, or lack of awareness, pointing to opportunities for clearer language documentation or redesign.
}

To identify the workflow entities that \changed{are most frequently changed, we rely on the} workflow paths extracted using \gawd (cf. RQ3).
These paths represent the hierarchical structure of the modified workflow entities (e.g., \entity{jobs.build.steps.3.if:always()}).
To uncover common patterns across workflows, we normalize these paths as follows: %
(1) we replace user-defined identifiers (such as job names) with generic placeholders (e.g., replacing \entity{jobs.build by jobs.\textless id\textgreater}); and
(2) we mask %
the position of elements within arrays or lists (e.g., replacing \entity{steps.3} by \entity{steps.\textless nr\textgreater}).
For example, the path \entity{jobs.build.steps.3.if} %
is normalized to \entity{jobs.\textless id\textgreater.steps.\textless nr\textgreater.if}. %
These normalisations enables the abstraction of change paths that are syntactically different but semantically equivalent across workflows.
This abstraction allows to group changes by their structural role within the workflow, such as jobs or steps, even when they differ in names or positions, enabling a better comparison across workflows.

Applying this normalization process to the \changed{changed workflows in our dataset,} we obtain 198 unique normalized paths, reflect all syntactic workflow entities that have been subject to changes.
We organize these paths into a tree structure that mirrors the nested key organization of YAML syntax: each tree node represents \changed{an} entity (e.g., \entity{jobs}, \entity{steps}, \entity{permissions}), and child nodes represent increasingly specific sub-entities.
The root of the tree represents the conceptual structure of a complete workflow file, aggregated across all workflows.
For each tree node, we compute the proportion of changes it accounts for by summing the counts from all corresponding subpaths.
This hierarchical view enables a structured analysis of how changes are distributed across different workflow entities.

\tab{tab:frequency_of_path_changes} presents the most frequently \changed{changed} paths within this hierarchy.
For clarity and interpretability, paths are grouped based on shared structural prefixes.
Each table entry thus represents a category of syntactic entities sharing a common prefix (e.g., \entity{jobs.\textless id\textgreater.steps.\textless nr\textgreater.if}), capturing all changes affecting any sub-element within that subtree of the configuration.
We only include the paths that cumulatively account for at least 0.2\% of all observed changes.

For each path, we report the proportion of changes associated with it relative to the total number of changes in the dataset.
This metric illustrates how frequently modifications occur at that specific location in the workflow structure.
Additionally, the table indicates the most common change types (\ie \textsf{M}odification, \textsf{A}ddition, or \textsf{R}emoval) observed at each path.
If a single change type  constitutes more than 50\% of the changes at that path, it is labeled as the \changed{primary change} type.
Otherwise, the two most frequent change types are reported.
This breakdown reveals which parts of the workflow tree structure are more actively changed and in which way.

\bigskip\noindent\changed{\textbf{Change concepts.}
Below, we report on the most frequent path changes in workflows, grouped by the workflow \emph{concepts} of RQ2 (cf. \tab{tab:change_catalog}).}

\begin{table*} [!t]
{\footnotesize
\begin{center}
\caption{\changed{Relative frequency of changed workflow paths, alongside their primary change type (\textsf{A}ddition, \textsf{M}odification, \textsf{R}emoval) and concept (using the shortcuts introduced in \tab{tab:change_catalog}.)}}
\label{tab:frequency_of_path_changes}
\begin{tabular}{l|r|l|l}
\textbf{entity} & \textbf{\% changes}  &  \textbf{primary change type} & \changed{\textbf{Concept}}\\
\toprule
\verb|jobs/<id>| & 91.05 & \verb|M| (62.0\%) & TS \\
\verb|  steps[<nr>]| & 70.10 & \verb|M| (69.2\%) & TS \\
\verb|    uses| & 22.40 & \verb|M| (99.8\%) & TC, TS \\
\verb|    run| & 12.99 & \verb|M| (99.7\%) & TS \\
\verb|    with| & 12.65 & \verb|M| (64.8\%) & TC \\
\verb|    name| & 3.27 & \verb|M| (87.5\%) & D \\
\verb|    env| & 2.29 & \verb|A| (48.4\%), \verb|R| (26.7\%) & RC \\
\verb|    if| & 2.09 & \verb|M| (55.2\%) & ER \\
\verb|    id| & 0.34 & \verb|M| (43.2\%), \verb|A| (40.7\%) & TC \\
\verb|    working-directory| & 0.28 & \verb|M| (50.2\%) & TC \\
\verb|  strategy| & 9.89 & \verb|A| (37.6\%), \verb|M| (34.6\%) & TS \\
\verb|    matrix| & 9.52 & \verb|A| (36.3\%), \verb|M| (35.8\%) & TS \\
\verb|      include| & 2.88 & \verb|M| (48.3\%), \verb|A| (31.4\%) & TS \\
\verb|      exclude| & 0.47 & \verb|A| (44.7\%), \verb|R| (29.4\%) & TS \\
\verb|  runs-on| & 1.47 & \verb|A| (2.2\%), \verb|M| (95.9\%) & RC \\
\verb|  needs| & 1.08 & \verb|A| (46.9\%), \verb|R| (28.1\%) & ER \\
\verb|  env| & 1.05 & \verb|A| (41.5\%), \verb|M| (33.1\%) & RC \\
\verb|  if| & 0.95 & \verb|M| (47.8\%), \verb|A| (38.7\%) & ER \\
\verb|  name| & 0.79 & \verb|M| (79.9\%) & D \\
\verb|  with| & 0.72 & \verb|M| (50.3\%) & TC \\
\verb|  container| & 0.43 & \verb|M| (82.2\%) & TS \\
\verb|  uses| & 0.36 & \verb|M| (99.5\%) & EC \\
\verb|  permissions| & 0.35 & \verb|A| (81.6\%) & RC \\
\verb|  timeout-minutes| & 0.25 & \verb|A| (58.2\%) & ER \\
\verb|  outputs| & 0.24 & \verb|A| (54.9\%) & TC \\
\verb|on| & 5.86 & \verb|A| (47.0\%), \verb|M| (28.1\%) & ET \\
\verb|  push| & 2.22 & \verb|A| (45.2\%), \verb|R| (29.2\%) & ET \\
\verb|  pull_request| & 1.62 & \verb|A| (49.6\%), \verb|R| (25.6\%) & ET \\
\verb|  schedule| & 0.54 & \verb|M| (73.8\%) & ET \\
\verb|  workflow_call| & 0.31 & \verb|A| (54.8\%) & ET \\
\verb|env| & 1.48 & \verb|M| (47.9\%), \verb|A| (33.5\%) & RC \\
\verb|name| & 0.81 & \verb|M| (96.8\%) & D \\
\verb|permissions| & 0.45 & \verb|A| (81.6\%) & RC \\
\verb|concurrency| & 0.29 & \verb|A| (68.7\%) & ER \\
\end{tabular}
\end{center}
}
\end{table*}

\emph{Task Configuration (TC).}
\changed{
The most frequently changed paths belong to the workflow \entity{jobs} %
\changed{(91.05\%)}, and more specifically their \entity{steps} %
\changed{(70.10\%)}.
Within steps, the most common changes involve the \entity{run} execution commands, and the
\entity{uses} and \entity{with} entities that reference Actions and their parameters.
Notably, nearly all observed changes to \entity{run} and \entity{uses} are \emph{modifications} (over 99\%).
For \entity{uses}, these modifications correspond to Action version updates to maintain workflow stability by keeping dependencies up to date.
Similarly, \entity{with} is frequently \emph{modified} (64.8\%), reflecting the need to adjust Action parameters as workflows evolve.
}

\emph{Task Specification (TS).} Frequently changed paths belonging to this concept relate to the matrix \entity{strategy} \changed{(9.89\%), primarily for the
\entity{matrix} and \entity{include} entities that tend to be changed through a mix of additions and modifications,}
whereas for \entity{exclude} removals are more common than modifications. This pattern indicates that while matrix configurations are frequently expanded or adjusted, exclusions are more often cleaned up or removed entirely.
\changed{Changes to the \entity{uses} entity that involve replacing a referenced Action (rather than updating its version) are also classified under the concept of \emph{Task Specification}, as they correspond to changing the workflow logic by altering what task is being executed rather than how it is configured.}

\emph{Execution Triggers (ET).} The most frequently changed \changed{workflow triggers are} \entity{push} and \entity{pull\_request}. \changed{They} are used to initiate workflows in response to making a pull request or commit to the repository.
For execution triggers, the primary change type is addition, followed by a smaller number of modifications.

\emph{Runtime Configuration (RC).} \changed{Environment variables (\entity{env}) in workflows, jobs or steps are changed relatively frequently, mostly through addition or modification.
The \entity{permissions}, which are used to give fine-grained control over what workflows and jobs can access,  are changed less frequently and mostly involved adding more fine-grained permissions.}

\emph{Documentation (D).} \changed{Relatively frequent modfications are observed to the \entity{name}  of workflows, jobs or steps, %
indicating that developers often change the names of these entities to maintain workflow readability.}

\emph{Execution Rules (ER).} \changed{The most frequently changed entity for this concept is \entity{if} (2.09\% at step level and 0.95\% at job level),} used for the conditional execution of jobs or steps.
Its primary change type is modification, reflecting adjustments made to match evolving execution conditions.

\begin{custombox}
  \changed{The top ten modified workflow paths account for nearly 80\% of all changes, mostly in the \entity{jobs}, particularly within \entity{steps}.
  The most frequently changed workflow entities (\entity{run}, \entity{used} and \entity{with}) map closely to the \emph{Task Configuration} concept.
  }
\end{custombox}

The remainder of this section dives deeper into the \changed{observed} evolution of specific workflow entities, and discusses the impact of such changes on the security and reusability of workflows.

\changed{
\smallskip\noindent\textbf{Usage of automated tools for workflow maintenance.}
\tab{tab:frequency_of_path_changes} revealed that \entity{uses} is the most frequently modified workflow entity. The manual analysis of RQ2 revealed that many of these changes are in fact version updates of reusable Actions or other versioned workflow entities.
We verified this on the full dataset of RQ4, observing a frequent use of automated dependency update tools, identified based on commit authors associated with bots. %
\github's built-in Dependabot was found to be used in \changed{69.2}\% of all workflow histories, and the third-party tool Renovate in \changed{21.0}\% of them.
They are primarily updating Action versions in the \entity{uses} entity, accounting for \changed{562,093} of the \changed{587,624} modifications involving version updates.

\noindent We also dived deeper in the kind of Action version updates proposed by these tools.
\gha supports four version formats:  fully specified version tags (\eg \textsf{v5.0.1}), major-only version tags (\eg \textsf{v5}), commit hashes (\eg \textsf{8f4b7f8\ldots}), and branches (\eg \textsf{main}). GitHub recommends commit hashes as the most secure option, since they allow for immutable version pinning. For version tags, \github recommends fully specified versions over major-only versions.
Our workflow history analysis revealed an encouraging trend towards more secure version usage, with a decrease in version tag usage by \changed{7,112} cases, and an increase in commit hashes by \changed{7,235}, nearly catching up in absolute numbers (239k vs. 242k).
The version tags also shifted towards the more secure option of fully specified versions with an increase of \changed{1,434} cases, compared to a decrease of \changed{667} cases for major-only tags.
Fully specified versions have now become as common as major-only tags (120K vs. 121K).
These results highlight the wide adoption of dependency automation tools, contributing to more secure workflow practices.}

\smallskip
\changed{
\noindent\textbf{Modifying workflow permissions.}
As another way to increase workflow security, \gha provides the \entity{permissions} mechanism to give fine-grained control over what workflows and jobs can access. Introduced in April 2021, they allow to enforce the principle of least privilege.
While changes to \entity{permissions} represent a small fraction of the observed workflow changes (cf. \tab{tab:frequency_of_path_changes}),
we observed an notable increase in the number of additions of permissions %
at workflow level (\changed{26,410}) as well as at job level (\changed{25,708}).
The most frequently added permission type at workflow level was \textsf{contents:read}, whereas a wider variety of permissions were observed at job level (\eg \textsf{contents:write}, \textsf{id-token:write}, \textsf{pull-requests:write}). These additive permission changes suggest that workflows start with default permissions that get refined gradually as the workflow evolves.
}

\smallskip
\changed{
\noindent\textbf{Tradeoff between control and reusability.}
The analysis in this section revealed that the most frequently changed workflow entities are \entity{steps},
mostly within the \entity{uses} and \entity{run} specifications.
During the manual analysis of RQ2 we observed multiple replacements of \entity{run} by \entity{uses} or vice versa.
Analysing the full dataset of RQ4, we observed \changed{9,827} cases of replacing \entity{run} by \entity{uses}, reflecting a shift toward exploiting reusable Actions. Doing so promotes workflow modularity and maintainability, but can come with increased security risks~\cite{10.1145/3643991.3644899}.
We also observed \changed{7,230} cases of replacing \entity{uses} by \entity{run}, reflecting a preference for using custom scripts to enable more fine-grained control, performance optimisation, and avoidance of untrusted dependencies.
}

\begin{custombox}
  \changed{An in-depth analysis of the syntactic change patterns in workflows reveals that:
   (i) automated tools play a major role in maintaining workflows, especially to keep versions up to date;
  (ii) there is a trend towards more precise version pinning, enhancing workflow reliability and reproducibility.
  (iii) workflows tend to improve upon security best practices (such as respecting the principle of least privilege and the use of commit hashes for Action versioning) even though there is still significant room for improvement;
  (iv) workflow maintainers make trade-offs between modularity (though the use of reusable Actions) and fine-grained control (through the reliance on custom scripts).
   }
\end{custombox}

\section{Discussion} \label{sec:discussion}

\changed{
In this section, we compare our findings with existing research on the evolution and maintenance of CI/CD configuration files.
Each subsection corresponds to one or more of our research questions, providing a focused discussion of how our results align with, extend, or differ from prior work.
\sect{subsec:workflow_file_evolution_and_maintenance_patterns} discusses workflow file evolution patterns (RQ1),
\sect{subsec:conceptual_changes_in_workflows} addresses conceptual changes observed in workflows (RQ2),
\sect{subsec:evolution_of_workflow_contents} examines the changes on the content of workflow files (RQ3 and RQ4) and compare those findings with prior studies.
}

\changed{
\subsection{CI/CD configuration file evolution patterns}
\label{subsec:workflow_file_evolution_and_maintenance_patterns}
This subsection compares the results of RQ1 to prior research.
Our analysis over 267K+ workflow change histories revealed that roughly 7\% of workflow files are modified each week, with a median of 82 days between successive updates. Modifications to workflow files clearly dominate over file additions, removals and renamings.
One out of four repositories add more than one workflow file (with a median of three) during their lifetime, and removals tend to occur shortly after their creation.
These results indicate that \actions workflows are undergoing a continued maintenance practice rather than a one-time setup.

Prior studies of CI/CD configuration evolution reported similar yet less pronounced trends.
Hilton \etal~\cite{hilton2016usage} investigated how often developers evolve their CI configurations by analysing the full history of \texttt{.travis.yml} files across 34,544 \github repositories.
They reported a median of 12 configuration changes per repository, with 25\% of them making five or fewer changes, suggesting that most teams follow a “set up once and adjust occasionally” pattern, while a minority frequently revise their CI setups.
In comparison, our analysis of GitHub Actions workflows revealed a more active evolution pattern, with a median of 22 workflow file changes per repository, with 14\% of them making five or fewer changes.

Zampetti \etal~\cite{zampetti2021ci} studied the evolution of 4,644 \github repositories, showing that \jenkins CI/CD configuration files typically change less often than production or test code but still undergo several updates throughout a repository's lifetime.
The median number of commits impacting \jenkins configuration files is 20, and 10\% of the studied repositories %
have more than 100 such commits.
In comparison, we found a median of 22 commits impacting \actions configuration files, with 10\% of the repositories having more than 121 commits.

The above observations suggest a slightly higher modification frequency for \actions workflows than for other types of CI/CD configuration files.
}

\changedTo{
We are not aware of any prior research having statistically studied the potential impact of LLM-based coding agents (such as Copilot) on the evolution patterns of CI/CD configuration files. This is unsurprising, given their relative recency and the many perils that need to be overcome to reliably detect their presence, use and evolution in GitHub repositories \cite{Robbes2026MSR}. The peril of \emph{partial observability} states that coding agents only leave partial traces of their activity (if at all). The perils of \emph{agent diversity and multiplicity} state that different agents may work quite differently and hence may leave traces in very different ways. Finally, the peril of \emph{high velocity} implies that the way agents function and are being used changes very rapidly.
Because of this, quantifying the impact of LLM-based agents on specific development practices is very challenging.
In this paper, we conducted a before-after analysis to try to detect whether their introduction had any significant impact on the change frequency of GitHub Actions workflow commits, but we observed only negligible effects at best. Further work at a more fine-grained level, based on the mitigation heuristics proposed in \cite{Robbes2026MSR}, would be needed to study any positive or negative impact of the introduction and use of specific LLM-based tools, either during the early phase or sustained lifetime of workflow configuration files.
}

\changed{
\subsection{Conceptual changes in workflows}
\label{subsec:conceptual_changes_in_workflows}

In RQ2 we manually analysed and classified 439 distinct workflow changes into seven concept categories. %
The results indicated that changes to \actions workflows predominantly focus on \textit{task specification} and \textit{task configuration}, mirroring developers' emphasis on refining execution logic and adjusting the parameters of reusable Action components.
Earlier studies reported similar high-level categories for other CI/CD services such as \travis and  \jenkins~\cite{beller2017oops,zampetti2020empirical,zampetti2021ci}.
For instance, the ``build logic'' category of~\cite{beller2017oops,zampetti2020empirical,zampetti2021ci} corresponds to our \textit{task specification} concept,
their ``environment configuration'' to our \textit{runtime configuration} concept,
and their ``documentation'' category to our \textit{documentation} concept.
This overlap demonstrates that \gha workflow maintenance is similar in nature to maintaining workflows or pipelines for other CI/CD services, though expressed through different syntactic constructs.
}

\changedTo{
We also analysed to what extent commits modify multiple workflow files simultaneously and what was the nature of these changes.
The majority of the identified cases (68 out of 95) consisted of the same kind of change being made to multiple workflows,  mostly for shallow routine maintenance-oriented activities.
While some of these activities (e.g. dependency updates) were automated via tools, we suspect that many other co-changes were actually performed manually.
This provides an opportunity for tool builders to better support cross-workflow maintenance beyond dependency automation.
For researchers, the presence of cross-workflow changes provides an opportunity to shift attention from studying isolated workflows to more ecosystem-wide studies of workflow changes, not only within individual commits, but even across different repositories. As an illustration of such research, Cardoen \etal \cite{Cardoen2026SANER} empirically analysed the prevalence of duplication within workflow files, but also across workflow files belonging to different repositories.
}

\changed{
\subsection{Evolution of workflow contents}
\label{subsec:evolution_of_workflow_contents}

RQ3 and RQ4 jointly investigated how the contents of individual workflow files evolve by analysing  what kinds of changes  (addition, removal, or modification) are made and which workflow entities are most frequently affected.
RQ3 identified 7.8M+ %
individual changes across all workflow histories, including 3.5M+ %
modifications, 1.5M+ %
additions, 887K+ %
removals,
and 48K+ %
renames.
Nearly four out of five changesets (78.98\%) consist of a single change type, most often \emph{modifications} (78.81\%).
This share grows over time, suggesting that workflow evolution is dominated by incremental refinements rather than major restructuring.
These quantitative results indicate that most workflows evolve through continuous adjustments rather than complete redesigns, mirroring earlier findings for other CI/CDs such as \travis and \jenkins~\cite{hilton2016usage,zampetti2021ci,ghaleb2024ci}.
\changedTo{We also statistically analysed possible effects of evolving technology (such as the introduction and use of LLM-based coding tools) on the frequency of changes to workflow entities, but we observed only \emph{negligible} differences in change frequency for \emph{modifications} to workflow entities.
As already mentioned in \sect{subsec:workflow_file_evolution_and_maintenance_patterns}, more fine-grained future research would be needed to study other possible effects, taking into account the many perils and pitfalls that come with such analysis \cite{Robbes2026MSR}.
}

RQ4 examined the distribution of changes across workflow entities, observing that nearly 80\% of all changes occur in the \entity{jobs} section, particularly within \entity{steps}.
Most of these correspond to \emph{task configuration} activities (\entity{uses}, \entity{run}, and \entity{with}).
A deeper examination of the \entity{uses} entity revealed distinct maintenance strategies.
Among 1.3M+ %
changes to this entity, 85.2\% correspond to Action version updates, while 14.8\% involve replacing the Action  entirely.
Such replacements arise when existing Actions become outdated, deprecated, or unmaintained, consistent with prior reports on ecosystem health and developer abandonment~\cite{constantinou2017empirical,avelino2019abandonment,kaur2022insights}.
Some of these Action replacements also aim to improve performance or optimise execution, similar to the optimisation opportunities identified by Bouzenia \etal~\cite{Bouzenia2024}.
For example, we found cases of workflows replacing \textsf{actions/download-artifact} with \textsf{actions/cache} to improve efficiency, or migrate from deprecated Actions (such as \textsf{ghaction-docker-buildx} which has been created and maintained by community) to officially supported ones by docker organisation.
Another type of change relates to replacing the \entity{uses} entity by  the \entity{run} entity to gain finer control and reduce dependency risks, or vice versa to increase workflow modularity. This duality reflects broader maintenance trade-offs between reusability and autonomy.

We observed a trend towards improved workflow security to reduce the \actions attack surface, through an increase in the use of \entity{permissions} to enforce the principle of least privilege, the extensive use of automated tools (such as Dependabot and Renovate) to keep versions up to date, and the increased use of commit-hash version pinning to secure Action versioning.
While prior research reported limited use of Action version pinning (only 1.6\% in~\cite{decan2022use}), our findings indicating a growing tendency  toward secure versioning practices since nearly half of all Action version updates employ commit hashes.
However, this observation is based on workflow changes rather than on the overall population of workflows, thus, the increased proportion of commit-hash updates could reflect more frequent updates within a subset of workflows rather than widespread adoption across all repositories.
Our observations about increased use of dependency automation tools for workflows aligns with prior findings that projects relying on such tools reduce their exposure to security issues and technical debt~\cite{Mohayeji2023Dependabot}.
With respect to the use of permissions, automated tools such as StepSecurity\footnote{\url{https://www.stepsecurity.io}} could help to further enforce the principle of least privilege, but remain underexploited.
These results align with broader observations on the evolving maturity of the \gha ecosystem~\cite{Decan2023JSS}, in which maintainers progressively balance reuse, security, and organizational consistency.

In summary, the analyses of RQ3 and RQ4 portray the continuous evolution of \actions workflows to satisfy diverse needs.
Workflows evolve through incremental modifications, focused on dependency management, configuration refinement, and security improvements, rather than large-scale redesign.
Future work could further contextualize these findings by linking workflow change histories with execution logs, such as those proposed by Moriconi \etal~\cite{Moriconi2025GHALogs}, and enriched metadata such as commit messages or issue references, enabling a more comprehensive understanding of how workflows evolve in response to runtime outcomes, errors, and developer intent.
}

\section{Threats to Validity} \label{sec:threats}

We follow the structure recommended by Wohlin \etal~\cite{wohlin2012experimentation} to discuss possible threats to validity of our research.

\textbf{Construct validity} concerns the relation between the theory behind the experiment and the observed findings.
The dataset of workflow files we used identified the presence of workflow files by detecting (valid) \yaml files within the designated \texttt{.github/workflows} directory. As a consequence, it may be the case that some workflows in this dataset were not actively used (i.e., never triggered for execution) or were not even intended to be used.
We expect this to be the case for a limited number of workflows, since the process of writing a workflow file and placing it in the appropriate directory is a deliberate action taken by developers.
Moreover, since changes to workflow files reflect developer effort and intent, the effort of maintaining workflow files remain valuable independently of whether the workflow is actively being used or not.

A second threat arises from our decision to exclude workflow histories containing invalid workflow files, as explained in Section~\ref{subsec:dataset}. We did so in order to ensure quality and syntactic correctness of the data under scrutiny.
This led to the exclusion of \changed{11.54}\% of all workflow files histories.
Those workflow histories may exhibit a different change patterns compared to valid workflow files, particularly in terms of debugging.
However, there is no evidence that this would be the case.

A third threat relates to the fact that we only considered the primary branch of each repository
when detecting workflow file changes, excluding changes occurring in parallel branches. While this prevented us from studying fine-grained changes in parallel branches, we argue that this is not a significant threat to our study. Indeed, the primary branch is typically the most active one, reflecting the development state of the repository. Additionally, changes made in parallel branches are usually merged back into the primary branch or discarded, meaning that the primary branch will eventually reflect all changes made in the repository.
The main bias related to this threat stems in how changes are merged from parallel branches to the primary branch: changes from parallel branches can be merged with a squashing strategy, implying that multiple distinct changes are combined into one large changeset, potentially altering the observed evolution patterns.
Unfortunately, there is no way to detect whether a squashing strategy was used or not, as the \git history does not retain information about the merging strategy~\cite{bird2009promises}.

\textbf{Internal validity} relates to the extent to which the study results are influenced by the experimental treatment or condition being studied.
One such treat is related to the \gawd tool we used for analyzing workflow file changes.
The tool can be parameterised in many ways to detect different types of changes in workflow files.
We relied on \gawd's default configuration for our analysis, but other settings might produce different results.
To mitigate this threat, we manually checked the results of \gawd on a small sample of workflow file changes to ensure that the default configuration was appropriate for our study.

\textbf{Conclusion validity} concerns the degree to which reasonable conclusions have been derived from our analysis.
The qualitative analysis in RQ2 was based on a manual analysis of a limited number of workflow files, potentially leading to an incomplete classification of change concepts.
We mitigated this threat by analysing workflow file changes until saturation was reached, ensuring no new change types were emerging.
Another potential threat to conclusion validity arises from the subjective interpretation of changes in workflow files. This subjectivity could lead to misclassification or biased analysis of the observed changes, potentially impacting the reliability of our findings.
To mitigate this risk, we followed the well-established framework method~\cite{spencer2004quality} to guarantee a structured and systematic qualitative analysis. Additionally, we involved multiple researchers in the process to cross-verify interpretations, thereby reducing the likelihood of individual biases and ensuring a more robust understanding of the data.
Finally, the qualitative findings from RQ2 were triangulated with quantitative results in RQ4, further supporting the validity of our conclusions.

\textbf{External validity} concerns the generalisability of the results beyond the specific data being analysed.
The dataset we relied on to conduct our study is based on active \github repositories (i.e., those that had at least one commit since \changed{October 2024}), that have at least 300 stars and 300 commits.
The rationale behind these criteria is to exclude repositories that are not representative of typical software development practices, such as abandoned, personal, or experimental repositories~\cite{Kalliamvakou2016}.
While we share the rationale behind these criteria, it is important to note that our findings may not be generalisable to smaller or less active repositories that may employ workflows for different purposes, such as publishing GitHub Pages or managing personal projects.
Consequently, the applicability of our conclusions to these contexts remains uncertain.
\section{Conclusion} \label{sec:conclusion}

This article presented a mixed-methods empirical study of how \github Actions workflows evolve over time.
We examined how workflows change throughout their lifetime, which conceptual areas are most affected, and which syntactic entities are most frequently modified.

The qualitative part of the analysis involved a manual investigation of  \changed{1,109} distinct changes across \changed{439} workflows. We clustered these changes along three change types (modification, addition and removal) into seven conceptual categories. Modifications were the most frequent, and the changes primarily belonged to the concepts of task specification and task configuration.
\changedTo{We also observed that commits that change multiple workflows are not uncommon, but mostly correspond to routine maintenance-oriented activities that make the same kind of change to multiple workflows.}

The quantitative part of the analysis relied on a large dataset of \changed{267,955} workflow histories, encompassing \changed{3,418,911} distinct workflow files across \changed{49,258} repositories over a period of \changed{six years}.
Around one out of four repositories added more than one workflow file during their lifetime. %
Workflow files were updated on average every \changed{159} days, with about 7\% %
of all workflows being modified every week.
\changedTo{The dominant change type was \emph{modifications} to the workflow contents. %
We also observed \emph{bursts} of consecutive workflow commits, reflecting iterative debugging and trial-and-error maintenance behaviour.
We analysed whether the workflow change frequency or burst behaviour differed between the period before and after major technological changes in GitHub (such as the emergence of LLM-based coding tools like Copilot), but could not find any  conclusive evidence.}

An automated analysis of workflow diffs revealed that the large majority %
of changesets consists of modifications as the only change type. %
Such modification-only changesets become increasingly dominant over time, suggesting that most projects change their workflow configurations iteratively. %
Delving into these workflow modifications, we observed that nearly \changed{80\%} of them are concentrated in only ten workflow paths.
The majority of them involve changes in the workflow jobs, and more specifically their steps. Such changes map directly to the concepts of task specification and task configuration.

Automated tools were found to play an important and growing role in workflow maintenance.
Dependency version update tools account for roughly one out of five workflow changesets.
Moreover, there is an increasing trend towards safer versioning practices %
and more fine-grained permission control. %

In summary, we have shown that \github Actions workflows are dynamically maintained software configuration components that evolve continuously through iterative, fine-grained maintenance.
Yet, several promising future directions  remain.
A key research opportunity is to study the co-evolution of workflow configurations and their execution behavior.
Workflow changes are likely driven by performance, reliability, or dependency issues.
Moreover, enhancing workflow functionality often requires iterative trial-and-error processes.
Empirical analyses of datasets combining workflow histories, execution logs, and commit messages could reveal deeper insights into the motivations and consequences of workflow evolution.
Another research avenue is to advance tool support for workflow maintenance beyond mere dependency management. Developers could benefit from intelligent tools to detect workflow smells, assist debugging and testing, detect and fix vulnerabilities, and recommend improvements tailored to repository context and maintainer needs.
Evaluating the effectiveness of these tools through empirical studies could further bridge the gap between workflow engineering and practical automation.
\changedTo{A final untapped opportunity consists of understanding and quantifying the fine-grained impact on workflow maintenance of the introduction and use of specific AI-based agents, either during the early phase or sustained lifetime of workflow configurations, considering the perils and mitigation heuristics proposed by Robbes~\etal~\cite{Robbes2026MSR}.}

\section*{Acknowledgments}

This work is supported by the ARC-21/25 UMONS3 Action de Recherche Concertée financée par le Ministère de la Communauté française - Direction générale de l’Enseignement non obligatoire et de la Recherche scientifique, and by the Fonds de la Recherche Scientifique - FNRS under grant numbers T.0149.22, F.4515.23 and J.0147.24.


%
%

\end{document}